\documentclass{bmcart}
\usepackage{amsthm,amsmath}
\usepackage{graphicx}
\usepackage{amsfonts} 
\usepackage[utf8]{inputenc} 
\usepackage{rotating}
\DeclareMathOperator*{\argmax}{arg\,max}
\definecolor{myyellow}{RGB}{200, 210, 0}
\definecolor{myolive}{RGB}{95, 180, 0}
\definecolor{mydarkgreen}{RGB}{0, 100, 100}

\newcommand{\supplementarysection}{%
	\setcounter{figure}{0}
	\let\oldthefigure\thefigure
	\renewcommand{\thefigure}{S\oldthefigure}
	\section{Supplementary section}
	\let\oldchapter\chapter
	\renewcommand{\chapter}{
		\let\thefigure\oldthefigure
		\let\chapter\oldchapter
		\oldchapter
	}
}

\startlocaldefs
\endlocaldefs

\begin{document}
	
	\begin{frontmatter}
		
		\begin{fmbox}
			\dochead{Research}
			
			
			\title{Ranking of Communities in Multiplex Spatiotemporal Models of Brain Dynamics}
			
			
			\author[
			addressref={aff2,aff1},                   
			corref={aff2},                       
			email={j.wilsenach@ucl.ac.uk}   
			]{\inits{J.B.W.}\fnm{James B.} \snm{Wilsenach}}
			\author[
			addressref={aff3},
			email={katie.warnaby@ndcn.ox.ac.uk}
			]{\inits{C.E.W.}\fnm{Catherine E.} \snm{Warnaby}}
			\author[
			addressref={aff1}, 
			email={deane@stats.ox.ac.uk}
			]{\inits{C.M.D.}\fnm{Charlotte M.} \snm{Deane}}
			\author[
			addressref={aff1,aff4}, 
			email={reinert@stats.ox.ac.uk}
			]{\inits{G.D.R.}\fnm{Gesine D.} \snm{Reinert}}
			
			\address[id=aff2]{%
				\orgdiv{Wellcome Centre for Human Neuroimaging},
				\orgname{Institute of Neurology, University College London},
				\city{London},
				\cny{UK}
			}
					\address[id=aff3]{%
			\orgdiv{Wellcome Centre for Integrative Neuroimaging},
			\orgname{Nuffield Department of Clinical Neurosciences, FMRIB Centre, University of Oxford},
			\city{Oxford},
			\cny{UK}
		}
						\address[id=aff1]{
				\orgdiv{Department of Statistics},             
				\orgname{University of Oxford},          
				\city{Oxford},                              
				\cny{UK}                                    
			}
					\address[id=aff4]{%
			\orgdiv{The Alan Turing Institute},
			\city{London},
			\cny{UK}
		}
			
			
		\end{fmbox}	
		\begin{abstractbox}
			
			\begin{abstract} 
				As a relatively new field, network neuroscience has tended to focus on aggregate behaviours of the brain averaged over many successive experiments or over long recordings in order to construct robust brain models. These models are limited in their ability to explain dynamic state changes in the brain which occurs spontaneously as a result of normal brain function. Hidden Markov Models (HMMs) trained on neuroimaging time series data have since arisen as a method to produce dynamical models that are easy to train but can be difficult to fully parametrise or analyse. We propose an interpretation of these neural HMMs as multiplex brain state graph models we term Hidden Markov Graph Models (HMGMs). This interpretation allows for dynamic brain activity to be analysed using the full repertoire of network analysis techniques. Furthermore, we propose a general method for selecting HMM hyperparameters in the absence of external data, based on the principle of maximum entropy, and use this to select the number of layers in the multiplex model. We produce a new tool for determining important communities of brain regions using a spatiotemporal random walk-based procedure that takes advantage of the underlying Markov structure of the model. Our analysis of real multi-subject fMRI data provides new results that corroborate the modular processing hypothesis of the brain at rest as well as contributing new evidence of functional overlap between and within dynamic brain state communities. Our analysis pipeline provides a way to characterise dynamic network activity of the brain under novel behaviours or conditions.
			\end{abstract}
			%
			
			
			
		\end{abstractbox}
		%
	\end{frontmatter}
	
	
	
	\section{Introduction}
	The brain activity of healthy subjects at rest is commonly used as a baseline against which a wide range of both pathological (e.g. dementia) and healthy (e.g. sleep) conditions are compared \cite{de2018comprehensive,mitra2017resting,pullon2020granger}. Often, activity under one condition is modelled as a single static pattern of activity, ignoring large scale dynamic shifts. However, neuroimaging researchers have begun to recognise that subjects move through a wide array of brain activity configurations even while relaxed or asleep \cite{vidaurre2016spectrally,karahanouglu2017dynamics,suk2016state}. A brain state is a configuration of brain activity evoked in response to a stimulus or to facilitate more complex responses \cite{brown2006brain}. Neuroimaging time series provide a way to observe these reconfigurations as spatial patterns of metabolic or electrophysiological activity, termed \textit{functional activity} \cite{papo2019gauging}. In order to generate these patterns, brain regions must coordinate through transfer of information. This exchange between brain regions defines the state's \textit{functional connectivity}. Functional activity can therefore be interpreted as a realisation from a brain state graph model which describes brain dynamics and the relationships between brain regions in the state \cite{bassett2017network}. This relates to models of the relationship between observed state and environment in which states are realisations of a so-called Markov blanket taking input from the environment to create an internal model of the external and internal environment \cite{hipolito2021markov,kirchhoff2018markov}. In these graph models, nodes are anatomically or functionally defined brain regions and edge strength is determined by the level of information shared between these regions (their functional connectivity). 
	\\*[4pt]
	The dynamics of communities of brain regions are of particular interest due to the important functional roles some communities play. Previous work has focused on deriving communities of brain regions using a number of methods including dynamic community detection \cite{martinet2020robust}. State space models have also been proposed that focus on the changing community structure within brain states from inferred functional connectivity \cite{ting2020detecting,liu2018global}. Our novel framework uses a Hidden Markov Model (HMM) approach  to construct a model, we term a Hidden Markov Graph Model (HMGM). This framework is fully unsupervised requiring no sliding window-based estimation or thresholding of the functional connectivity, and no prior assumptions about the number of states or embedding dimension.
	\\*[4pt]
	We analyse brain state dynamics as a multiplex graph with modular (community) structure at both the temporal (state switching dynamics) and spatial (brain region communication) levels. In order to differentiate the temporal communities of states and  the less functionally relevant spatial communities from the most relevant we use the term \textit{network}. Network here is used exclusively to refer to modular subgraphs of coordinated brain regions within a state that are functionally important (rather than being synonymous with the term graph). These brain networks form the basis of our understanding of the functional connectivity pathways within the brain and are integral to our understanding of the role of changing brain configurations in wakefulness and beyond \cite{rosazza2011resting}. 
	\\*[4pt]
	We have developed a method based on the HMGM framework to identify the importance of possible brain networks using random walks to ascribe to each module in each state an importance or $T$-score based on their functional connectivity and co-activation. Notably, the method does not apply random walk information to partition the graph but rather to determine the relative importance of communities within a partition \cite{rosvall2008maps}. Our method provides a means to characterise dynamic functional activity under novel conditions or behaviours. As a proof of principle, we apply our pipeline to neuroimaging data from subjects at rest and provide new evidence for both modular and nested functional activity in the awake brain. 
	\subsection{Static Brain State Models}
\begin{figure}[!h]
		\centering
		\includegraphics[width=0.97\textwidth, height=0.4\textheight]{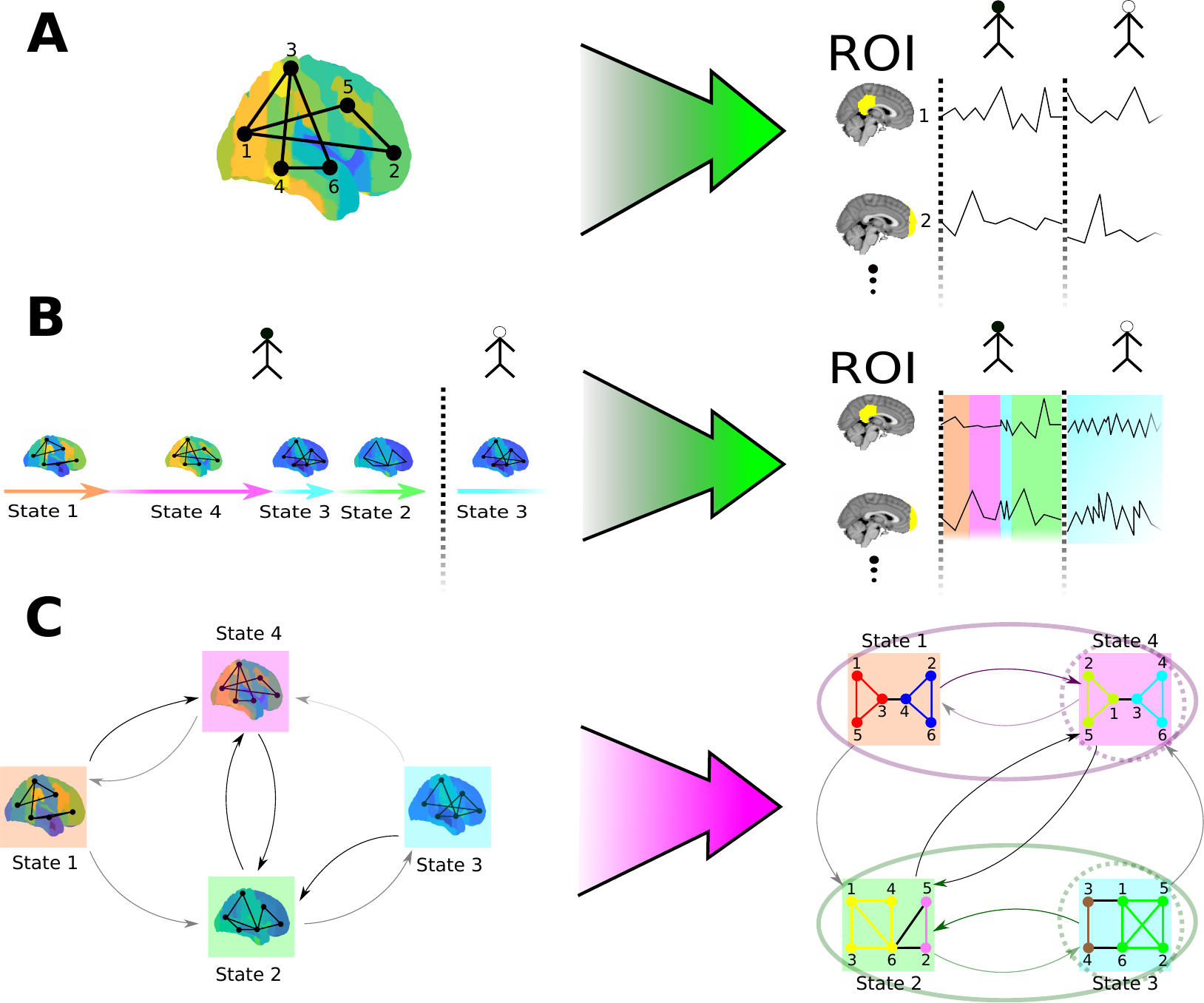}
		\caption{This figure shows how brain activity can be modelled as being generated by  a system of either static (A) or dynamic (B) states, and, in particular, how such a dynamic brain state model can be interpreted as a multiplex network with modular structure (C). In \textbf{(A)} a static pattern (left) of functional activity (colour of functional activity map) and connectivity (edges between regions) is observed (green arrow) as a stationary multi-ROI multi-subject time series (right) in which each dimension is the activity observed for a particular Region of Interest (ROI) in each subject (separated by a dotted line). \textbf{(B)} Shows state dynamics for a multi-subject system with multiple states (left). In this system each state is represented by a colour and arrow length indicates its duration in time. This is observed as a multivariate time series composed of weakly stationary segments (right). Segment colour indicates the state that generated it. In \textbf{(C)} we use temporal relationships between states to represent the system as a dynamic multiplex graph. This system is decomposed (purple arrow) into its essential temporal (coloured ellipses) and spatial modules or functional \textit{networks} (coloured subgraphs). Dashed circles around states show state hubs, important states in each community which are central to the dynamics and facilitation of brain activity across subjects.  \label{fig:multilayer}}
	\end{figure}

	In the simplest brain state models (see Figure \ref{fig:multilayer}A), functional activity arises as noisy realisations of a single static brain state. Considerable progress has been made using this static framework to characterise the vast repertoire of activity patterns observed during wakefulness. Models using both weighted and
	unweighted graph structures derived from Independent and Principal Component Analysis (ICA and PCA respectively) have revealed key modules within the brain across a wide range of conditions. These include both behavioural and task-based conditions (sensory, motor etc.) and resting state conditions in the absence of direct stimulation \cite{calhoun2012multisubject,smith2009correspondence,kokkonen2009preoperative,samann2011development,calhoun2004method}. Recent results from both electrophysiological data derived from Electroencephalography (EEG) and Blood Oxygen Dependent (BOLD) data derived from functional MRI (fMRI), suggest that weighted network models produce more reliably reproducible and robust results than do binarised network models \cite{jalili2016functional,ran2020reproducibility,smith2017accounting}.
	\\*[4pt]
	Studies using static models have helped neuroscientists to build up vast libraries of associations between cognitive functions and specific brain regions \cite{poldrack2011cognitive}. However, the static approach makes it difficult to account for inter-subject variability as well as dynamic changes in state that occur in time as different cognitive and functional demands are placed on the brain \cite{michael2014preserving}. These demands result in activity in one moment that is often functionally incompatible with activity in the the next, driving the need for dynamic approaches to brain state modelling \cite{sridharan2008critical}.
	\subsection{Dynamic Brain State Models \label{dynstate}}
	Moving window-based approaches produce a series of snapshots of the activity pattern of the brain. Although these methods have proved incredibly useful in understanding changing brain state, they are limited in their ability to reliably detect changes in functional connectivity between regions over time \cite{hindriks2016can}. By contrast, state space models (Figure \ref{fig:multilayer}B) and in particular Hidden Markov Models (HMMs) \cite{vidaurre2017brain,chen2016spatiotemporal}, have arisen as an alternative to the sliding window approach and use a number of simplifying assumptions to improve on these models' tractability and specifiability \cite{suk2016state}. More recently, dynamic community detection methods have been proposed which capture many of the same features as dynamic state space models, however these methods often still rely on sliding window approximations of functional connectivity to construct a series of dynamic networks \cite{martinet2020robust,liu2018global}.
	\\*[4pt]
	The chief underlying assumption of HMMs is that brain dynamics can be parametrised by a finite state, positive recurrent, Markov process where functional activity and connectivity is determined by an observation model, typically a multivariate normal distribution \cite{vidaurre2016spectrally,ting2020detecting}. In these models, dynamic switching between states can be interpreted as a temporal graph of probable state transitions (Figure \ref{fig:multilayer}C). The full model can thus be interpreted as a nested, or multiplex graph in which the layers are brain states (with brain regions as nodes) and the interlayer directed edges are transition probabilities between state layers.
	\subsection{Novel Multiplex Approach \label{novelmult}}
	A state characterises a pattern of activity across the whole brain at a given time; however, it is most often characterised in terms of just a few key subgraphs of interacting brain regions (see Figure \ref{fig:multilayer}C) \cite{ting2020detecting,liu2018global}. Much progress has been made to characterise the vast repertoire of activity patterns observed during resting states and task performance. These enquiries have given rise to a number of re-occurring and important networks, associated with a wide range of brain functions and behaviours \cite{shulman9others,biswal1995functional,menon2011large}. The most prevalent and widely characterised of these are the so-called \textit{resting state networks}, termed the Default-Mode (DMN), Salience (SN) and Central Executive (CEN) Networks as well as those active during sensory and motor tasks including: the sensorimotor, visual and auditory networks \cite{ryali2016temporal}. The mechanisms underlying these networks are interdependent with recruitment of one network often necessitating the further recruitment of other networks \cite{karahanouglu2017dynamics}. Conversely, some networks are known to be largely mutually antagonistic in activity, with DMN and SN activity generally being anticorrelated with sensorimotor-like activity in resting wakefulness \cite{vidaurre2018discovering}. 
	\\*[4pt]
	Although state space modelling of brain dynamics is a relatively young field, one key finding has been the multi-scale modularity of brain states. In particular, Louvain modularity-based community detection applied to the temporal graph of state transitions has shown that states are organised modularly into communities under a variety of conscious conditions including resting wakefulness and sleep \cite{vidaurre2017brain,stevner2019discovery}.
	\\*[4pt]
	In order to construct a set of plausible brain states models we train a number of HMMs with different numbers of states on resting state data. We then utilise our novel cross-validated maximum entropy procedure, based on the maximum entropy principle, to select the HMM that best generalises across subjects \cite{jaynes1957information}. We convert the selected HMM into a dynamic graph model by transforming the state covariance matrices into weighted, directed graphs based on the regional correlations within each state and node attributes given by the state mean activity. The intralayer network which we term the Markov Information Matrix of the states is motivated by an interpretation of brain states as realisations of an underlying Markov blanket or network as in \cite{hipolito2021markov}.
	\subsection{Ranking the Importance of Networks within the Brain}
	We perform two-level Louvain community detection to discover important communities of brain states (temporal communities) and brain regions within a state (spatial communities). We use community centrality statistics to identify the hub states of key activity in each network. Within each hub state we look at spatial community structure to determine the key actors in the dynamics of the model that may be important to the overall dynamics of wakefulness across subjects. 
	\\*[4pt]
	Random walks provide an effective way to construct representative samples from a graph in a way that preserves local structure \cite{dupont2006relevant,leskovec2006sampling}. In complex interdependent data sets random walk sampling can be used to remove baseline levels of interdependence and discern the most robust relationships in a one dimensional model, by conditioning out local inhomogeneity in noisy activity \cite{luecken2018commwalker}. Here, we extend this principle to network sampling across two dimensions, space and time. Our method is based on a non-parametric random walk statistic that combines a temporal walk between layers with a spatial walk between regions. We use random walks to sample plausible patterns of functional network activity from the local functional activity background. We then use the samples as a benchmark against which to score functional coordination in our spatial communities. This statistical score, termed the $T$-score, is simple to compute given the graph model and putative network and is inspired by a similar method for analysing large, complex protein graphs with metalayer information \cite{luecken2018commwalker}.
	\\*[4pt]
	Our method allows us to determine which spatial communities are highly co-activated or inactivated relative to the expected dynamics across states in that brain area, providing a generalisable procedure to determine functionally relevant brain state communities. Our within state community functional associations largely agree with macroscopic analysis of the state functional activity maps, but provide an additional layer of information in the form of networks that provide clarification and depth to our understanding of brain states at the mesoscale.
	\subsection{Metatextual and Network Analysis of Brain State Models}
	We use the powerful metanalysis tool, \textit{Neurosynth} \cite{yarkoni2011large}, to determine functional associations between each brain state, it's most important networks and important functional terms from the literature. \textit{Neurosynth} provides scores based on either correlations between brain images and the occurrence of a predefined set of terms in the literature or, in conjunction with the \textit{NIMARE} package \cite{NiMARE53:online}, a posteriori probabilities of  associations between the image and an exhaustive list of literature terms. Using these tools and images derived from our brain states, termed functional activity maps, we provide evidence to corroborate the modular processing hypothesis in resting wakefulness \cite{reichardt2006statistical}. Key to our findings is that the states associated with resting state networks tend to self-associate while being anticorrelated with sensorimotor associated states.
	%
	\section{Methods}
	\begin{figure}[!h]
		\centering
		\resizebox{105mm}{!}{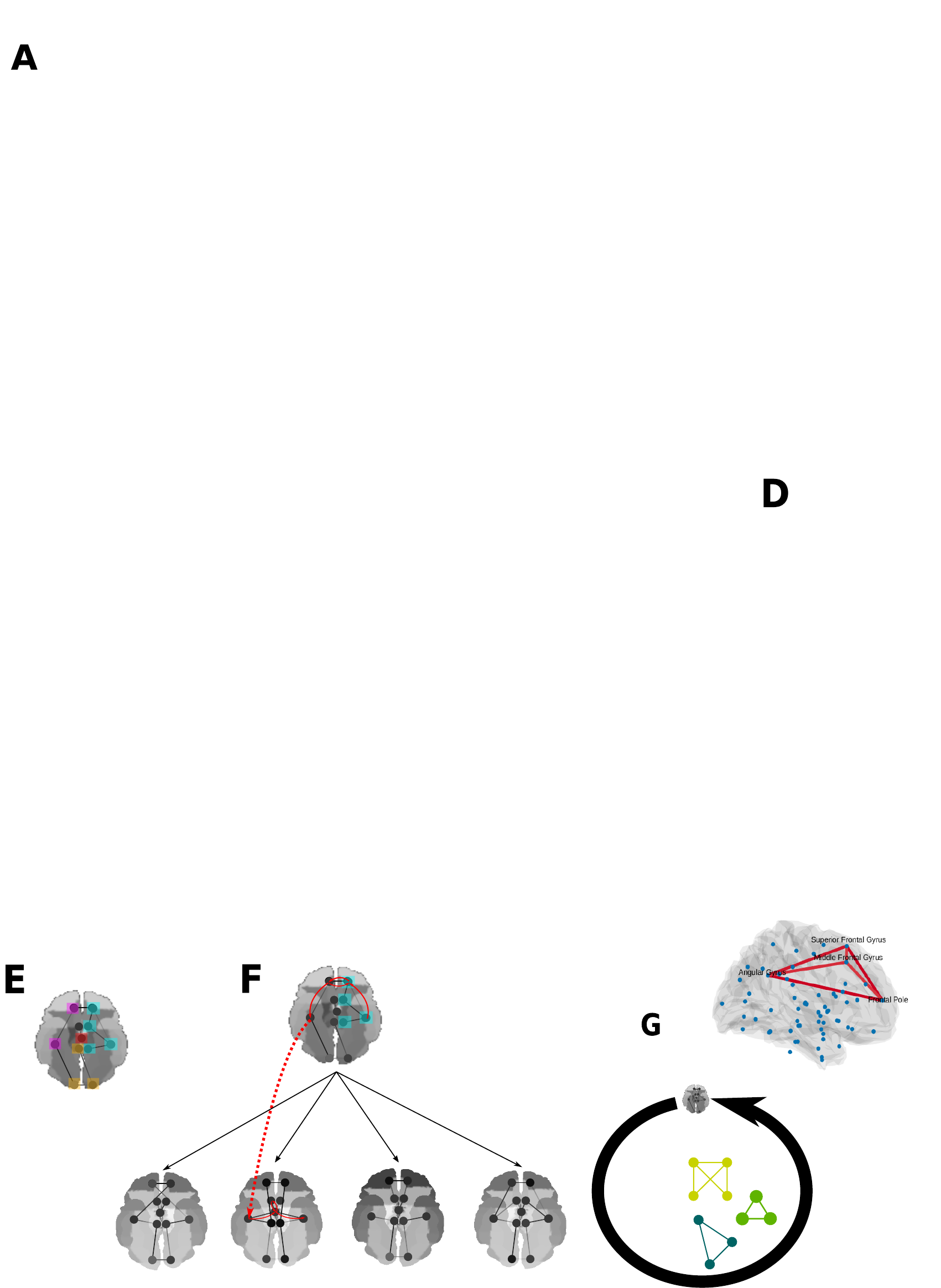}
		\caption{Flow diagram of the graph modelling and analysis pipeline. Following preprocessing of the fMRI data we obtain multivariate regional brain activity time series for all $N$ subjects. Variational Bayes inference is then used to train HMMs (using the \textit{HMM-MAR} package \cite{vidaurre2016spectrally}). \textbf{(A)} is a sketch of an HMM fitted to the $X_{n,t}$ data for subject $n$ and time point $t$. Each hidden brain state $S_{n,t}$  $\Sigma(S_{n,t})$ has mean activity $\mu(S_{n,t})$ and covariance $\Sigma(S_{n,t})$ (after backprojection). State change from $S_{n,t}$ is determined by the transition matrix $P$. \textbf{(B)} The number of hidden states, $K$, is determined using mean subjectwise cross-validated maximum entropy, which is calculated over the fractional occupancies, $\kappa_{s,n,k}$ for each subject-state pair up to $K$ states. \textbf{(C)} Adjacency matrix of the interlayer temporal directed transition graph determined by the Markov transition matrix of the HMM, with temporal communities in red along the diagonal. \textbf{(D)} Each state itself can be considered a layer with edges relating brain regions by their correlation in activity derived from their modelled covariance $\Sigma(s)$, with node weights (regional mean activity) determined by $\mu(s)$. Of the states, some are highly connected state hubs, $h(U)$, belonging to a temporal community $U$ (red shading). \textbf{(E)} Each hub state (layer) $h(U)$ is analysed and internal spatial communities are determined. \textbf{(F)} Internal communities are ranked according to their level of coherent brain activity compared to many repeated random walk samples from the multiplex model. \textbf{(G)} The results of ranking summarised by the community $T$-score. High $T$-score corresponds to a higher than expected level of community coherent activity when compared to the rest of the multiplex graph in this brain area. We propose functions for highly ranked communities by mapping these regions onto a 3D functional activity map and compared them to maps and terms drawn from the neuroscience literature with \textit{NeuroSynth}. \label{fig:flow}}
	\end{figure}
	In the following sections, Sections \ref{prep} and \ref{msel}, we explain the preprocessing of the data and define the state space (HMM) model and novel model selection criterion. We will see that each brain state $s$ can be thought of as a pattern of activity represented by a weighted graph $G(s)=\{V,a(s),W(s)\}$ in which each node is a brain region $x\in V$, (with $|V|=D$ nodes), each with a level of functional activity $a(s)^x$ attributed to $x$. Similarly, each edge in $G(s)$ is weighted by $W(s)^{x,y}\in W(s)$ the level of information flow from region $x$ to region $y\in V$ (edge absence is represented by $W(s)^{x,y}=0$), with $W(s)^{x,y}\ne W(s)^{y,x}$ in general. 
	\\*[4pt]
	As we shall show, Hidden Markov Modelling with our new model selection method, provides a means to construct a dynamic state space model from multi-subject fMRI time series data in a data driven way. We use inter-regional correlations to determine the state graphs and use the temporal relationships between states to determine the directed interlayer edges (see Figure \ref{fig:multilayer}C). Lastly, in Sections \ref{clust} to \ref{neurosynth} we set out methods to explore the spatiotemporal modular and functional structure of these multiplex brain state models.
	\subsection{Acquisition and Pre-processing of fMRI Data for HMM Modelling \label{prep}}
	Ten minutes of whole brain fMRI activity were recorded separately for each of $N=15$ wakeful subjects (with eyes closed) as part of a previous study \cite{mhuircheartaigh2013slow}. The brain volumes produced by the scanner were aligned to the MNI152 standard brain template \cite{fonov2011unbiased}. This resulted in a high dimensional time series of each subject's fMRI (BOLD) signal for each voxel, with a temporal resolution of 3s and a spatial resolution of 2mm$^3$ \cite{woolrich2009bayesian}.
	\\*[4pt]
	Recordings were collected separately from each subject. Of the 200 volumes recorded per subject (each time point is one volume), four dummy volumes were removed to exclude any non-steady-state magnetisation effects. This was followed by motion correction with MCFLIRT (Motion Correction FMRIB's Linear Image Registration Tool), spatial smoothing using a Gaussian kernel of 5mm full width half-maximum, global intensity normalisation, and temporal high-pass filtering with a cutoff of 0.02Hz to remove low frequency scanner drift. Automated removal of non-brain tissue was initially performed before statistical analysis using BET (Brain Extraction Tool), with further manual correction in FSLview. Further spatiotemporal artefact removal was carried out by independent component in FSL melodic \cite{woolrich2009bayesian}. 
	\\*[4pt]
	We selected regions of interest in our study based on the Harvard-Oxford (HO) probabilistic cortical and subcortical brain parcellations, which assigns to each voxel a probability for each brain region. We assign each voxel a unique region identity according to the maximum probability across regions in the HO parcellation.  Excluding white matter regions the resulting parcellation of 63 Regions of Interest (ROIs) includes 48 cortical and 15 subcortical brain regions \cite{caviness1996mri,makris2006decreased}. ROI time series were calculated using the ROI spatial mean BOLD signal at each time point. This results in a $D=63$ dimensional time series with $T=196$ time points per subject. Each of the $D$ constituent ROI time series were temporal mean subtracted and normalised by the standard deviation. 
	\\*[4pt]
	Model fitting presents two challenges, the first is that the time taken to fit the model scales with parametric complexity, and the second is that a poorly parametrised model may lead to overfitting or underfitting. To address these challenges, dimensionality reduction by principal components of the original $D$ dimensional time series was performed to reduce parametric complexity while also reducing overall noise. This approach is justified by the generally low embedding dimension of most real world data, including neuroimaging data \cite{ma2018randomly,shen2008low}. In order to balance dimensionality reduction and retention of signal, Parallel Analysis is used (see Supplementary Information Section \ref{supp:pa}) to obtain a $D\times d$ eigenmatrix $A$ of the first $d<D$ eigenvectors \cite{horn1965rationale}. This method assumes roughly linear separability of uncorrelated noise from signal, but has been shown to outperform a number of methods, including maximum likelihood estimation, in simulation \cite{humphreys1975investigation}. The reduced $d$ dimensional time series $\{X_{n,t}^*\}_{t\in \mathbb{N}_T}$ is then inputted to train a noise reduced HMM model of the data.
	\subsection{Model Specification and Generalisability \label{msel}}
	We use the \textit{HMM-MAR} package to train HMMs with multivariate normal observations by Variational Bayes \cite{vidaurre2016spectrally}, whilst separating the data by subject into distinct trials of length $T$. For further details on model fitting see \cite{vidaurre2016spectrally}. Figure \ref{fig:flow}A shows how observations of the fMRI BOLD signal at each time point are modelled across subjects. Dynamics for each subject are modelled and fitted using a shared set of states $\mathcal{S}$ with finite ${S}=\{1,2,...,K\}$ and Markov transition matrix $P$.
	\\*[4pt]
	We give a brief overview of HMM dynamics. We note that a key parameter, for these dynamics, the number of brain states, $K=|\mathcal{S}|$, that best generalises these dynamics across subjects is unknown. Consequently, we introduce a novel framework for selecting $K$ based on an information theoretic criterion that maximises generalisability by maximising entropy of the state dynamics across subjects.
	\\*[4pt]
	In each HMM state trajectory, the initial state of each subject's trial is selected independently at random. Under the Markov assumption of the model the resulting subject-specific state dynamics are assumed independent realisations of the same stochastic process, $S_{n,t}$. For $t>1$, $S_{n,t}$ is conditionally dependent on the previous time step $S_{n,t-1}$ so that
	\begin{align}
	Pr(S_{n,t}=s|S_{n,t-1}=s',\mathcal{M})=P_{s,s'},
	\end{align}
	for $s'\in \mathcal{S}$. Each brain state $s\in \mathcal{S}$ is associated with an observation model $O(s)\sim MVN(\mu^*(s),\Sigma^*(s))$. The $O(S_{n,t})$ model the row dimensionally reduced brain data $X_{n,t}^*$. In order to obtain the full model, the reduced model is then back-projected into $D$ dimensional brain region space (see Equation \eqref{eq:back}).
	\subsection{Novel Model Selection Criterion Based on Fractional Occupancy}
	The Markov chain defined by $P$ and any given initial state $s_0\in \mathcal{S}$, has a unique stationary distribution $\pi_s$ that is independent of $s_0$ assuming the chain is irreducible and the states are positive recurrent. The probability $\pi_s$ is the long run probability of the re-occurrence of state $s$. Selection of the number of these hidden states is carried out by cross-validated entropy maximisation over the related fractional occupancy distribution. The fractional occupancy distribution $\kappa$ is defined by subject $n$ for each state $s$ and given by
	\begin{align*}
	\kappa(s,n|\mathcal{M},X) = \frac{1}{T}\sum\limits_{t=1}^T Pr(S_{n,t}=s|\mathcal{M},X)
	\end{align*}
	where $P(S_{n,t}=s|\mathcal{M},X)$ is the posterior probability of state $s$ occurring at time $t$ given the model $\mathcal{M}$ and data $X$. The fractional is the probability of finding subject $n$ in $s$ over the entire trial of length $T$. The distribution $\kappa$ for subject $n$ is related to the stationary distribution $\pi_s$ by the well-known limit 
	\begin{align*}
	\kappa(s,n|\mathcal{M},X) \xrightarrow[\infty]{T} \pi_s.
	\end{align*}
	That is to say that $\kappa$ asymptotically approximates the long run average state dynamics of the model as trial length increases. Knowing this, our goal is to select the model whose fractional occupancy maximises the entropy pooled across subjects by maximising the objective function
	\begin{align*}
	H(k) = - \sum\limits_{n=1}^N\kappa(s,n|\mathcal{M}(n,k),X_n)\log[\kappa(s,n|\mathcal{M}(n,k),X_n)]
	\end{align*}
	where the model $\mathcal{M}(n,k)$ is the model trained using all trials except the data from subject $n$ assuming $k$ hidden states, and $X_n$ is the trial data from subject $n$ (see Figure \ref{fig:flow}B).
	\\*[4pt]
	By selecting the initial number of states $K=\argmax H(k)$, we appeal to the information theoretic principle of maximum entropy which states that the model which maximises the uncertainty over the data tends to be the one that best approximates the true data distribution \cite{jaynes1957information}. More specifically, our goal is to obtain a set of states with similar uncertainty about subject behaviour over the course of the experiment. We shall see in Section \ref{msel_res} that the goal of state-subject uncertainty maximisation relates closely to that of optimal model selection. We note that to the best of our knowledge this is the first application of such a subject-specific entropic criterion in state space model selection.
	\subsection{The State Markov Information Graph \label{mim}}
	First model parameters $\mu^*(s)$ and $\Sigma^*(s)$ for state $s$ from the HMM model $\mathcal{M}$ are backprojected using the transpose eigenmatrix $A$ to obtain a model in $D$ dimensional brain space so that the full $D$ dimensional model has mean $\mu(s)$ and variance $\Sigma(s)$ defined over the ROIs and given by
	\begin{align}
	\Sigma(s)=A\Sigma^*(s)A^T \qquad \textrm{and} \qquad \mu(s)=\mu^*(s)A^T.\label{eq:back}
	\end{align}
	Using the full model, each state $s$ has normally distributed observations with mean $\mu(s)^x$ and covariance $\Sigma(s)^{x,y}$, for brain regions $x,y \in V$. We use these to define a graph $G(s)=(V,a(s),W(s))$ over the set of $R$ brain regions, node weights $a(s)$ and edge weights $W(s)$, which we take to be a proxy for the information flow between regions. More specifically, we estimate the weights $W(s)$ by the correlation matrix $|\rho(s)|$, as derived from the state covariance matrix $\Sigma(s)$.
	\\*[4pt]
	Here, $a(s)^x=\mu(s)^x$ are the mean regional functional activity at brain region $x$ in $s$. The weighted edge (directed information flow) from regions $x$ to $y$ are
	\begin{align}
	W(s)^{x,y} = \frac{|\rho(s)^{x,y}|}{\sum\limits_{z=1}^D|\rho(s)^{x,z}|}.\label{markovFC}
	\end{align}
	The resulting edge weights matrix $W(s)$, defines a Markov transition matrix, a model of information flow between brain regions in state $s$ in which information flow between $x$ and $y$ is defined both into $x$ from $y$, $W(s)^{x,y}$ and out of $x$ to $y$, $W(s)^{y,x}$. Note this defines a potentially asymmetric and directed graph with edges (information flow) both into and out of $x$. The rationale for using such a Markov transition matrix to define edge weights is to convert the entire network into a dynamic Markov graph in which information is propagated probabilistically both in time and space. This is useful in particular in Section \ref{neurosynth}.
	\subsection{Louvain and Hierarchical Temporal Clustering \label{clust}}
	We perform Louvain modularity detection on the directed Markov transition and information graphs \cite{blondel2008fast}. Suppose $G=(V,E,W)$ is a potentially directed and weighted graph with vertex set $V$, edge set $E$ and weight matrix $W$. The Louvain algorithm involves the greedy optimisation of an objective function $Q(\mathcal{U})$, termed the modularity score for $\mathcal{U}$ a partition of $V$ (see Supplementary Information, Section \ref{supp:mod}) \cite{girvan2002community,newman2006modularity}. The algorithm allows for a resolution parameter $\gamma$ which determines the relative size of communities and goes to one as $\gamma \rightarrow \infty$ \cite{lambiotte2008laplacian}. 
	\\*[4pt]
	We use a form of the Louvain optimisation algorithm originally designed for undirected networks but complement this with a version of the modularity $Q(\mathcal{U})$ which has been adapted for directed networks in \cite{nicosia2009extending,leicht2008community}. In order to assess the validity of this approach, a rough measure of the degree of symmetry in a weight matrix $W$ can be given by the fraction of the energy of the adjacency matrix (as measured by the Frobenius norm) that is contributed by the symmetric part, $\textrm{Sym}(W)$ (see Supplementary Information, Section \ref{supp:asym}) \cite{aggarwal2020linear}. 
	\\[4pt] 
	In the case of temporal communities, we determine the significance of the community partitioning by comparing $Q(\mathcal{U})$ to an empirical distribution composed of modularity scores from 10,000 partitions constructed by random permutation of the community labels. In addition, in order to examine the state-subject relationships directly, we perform agglomerative hierarchical linkage clustering based on correlation in fractional occupancy $\kappa$ using Ward's method \cite{ward1963hierarchical}. 
	\subsection{Community Hub Selection}
	State hubs are the states most central to the dynamics of the model and facilitate the switching dynamics within each community. These are selected by maximising the community centrality z-score, $z(s)$, for each community $U\subset \mathcal{S}$ \cite{guimera2005functional,shine2016dynamics}. This score measures the within community degree centrality of a node relative to the mean community connectivity (see Supplementary Information \ref{supp:centrality}). Hubs are then analysed for their community structure, using the same Louvain algorithm as in \ref{clust} but this time on the directed brain state graph $G(s)$.
	\subsection{Identifying Functionally Important Spatial Communities \label{hubs}}
	Not all detected communities are as relevant to a state's functional role as others. Performance of these roles requires both functional activation and coordination of brain regions. To discern which communities are the most functionally cohesive, we rank communities by comparing to samples of regional activity from the full multiplex graph model (see Figure \ref{fig:flow}C and D).  We used random walks to sample plausible patterns of functional network activity and employ them as a benchmark against which to measure the level of coordination within spatial communities. Controlling for the local level of background activity in space and time allows for a more representative indication of functional cohesion within brain networks identified by community detection than naive comparison of communities by community mean functional activity.
	\\*[4pt] 
	We introduce to neuroimaging the Functional Homogeneity, $FH$, as our community coherence measure, a statistic derived from the mean activity $\mu(s)$ and $\Sigma(s)$ that is high when the community mean activity is most in agreement with the directions of maximum community functional connectivity and low otherwise. It is a measure of the alignment between the two key features of spatial communities, their level of shared information and activation. This measure is well suited for neuroimaging data, and is well established in computer vision and image classification where it is known as the covariance metric and measures the agreement between and within image classes \cite{li2019distribution}. The $FH$ for a community $C$ in a state $s$ is
	\begin{align}
	FH(s,C) = {\mu(s)^C}^T\Sigma(s)^C \mu(s)^C,
	\end{align}
	where the superscript $C$ refers to the submatrix given by removal of all rows and columns not corresponding to regions in community $C$. This metric is key to the community ranking procedure which follows a six step process:
	\begin{enumerate}
		\item Given a community $C\subset V$ in state  $G(s)$ we calculate $FH(s,C)$.
		\item Sample a state $s'$ from the stationary distribution $\pi$.
		\item Select a region $x\in C$ and sample $|C|$ nodes from $G(s')$ starting at $x\in V$ in $G(s')$.
		\item Repeat steps 2 and 3 to construct a representative sample of paired states and brain regions $(s_1,C_1),(s_2,C_2)...,(s_L,C_L)$
		\item Calculate the $T$-score for functional cohesiveness of a subgraph
		\begin{align*}
		T(s,C) = \frac{1}{L}\sum\limits_{l=1}^L I[FH(s,C)>FH(s_l,C_l)]
		\end{align*}
		where $I$ is the standard indicator function and rank the communities in $s$ by decreasing $T$-score.
		\item Determine whether the community represents a correlated or anticorrelated brain subgraph by the sign of $E_C[\mu(s)]=\sum_{x\in C} \mu(s)^x$.
	\end{enumerate}
	The $T$-scores of all the communities in a specific state can then be used to order the states in terms of which are most likely to contribute to the functional cohesion of the state. Note that $T(s,C)$ is a score between zero and one, with one implying that the community $C$ is much more functionally cohesive than other comparable brain subgraphs in space and time. $T$-scores are not designed to be compared across states. These steps are summarised by steps E to F in Figure \ref{fig:flow}.
	\subsection{Analysis of States and Communities with \textit{NeuroSynth} \label{neurosynth}}
	\textit{NeuroSynth} is a meta-analysis tool that takes in 3D images of brain activity (termed functional activity maps) in MNI152 standard space and returns a scored association (based on the Pearson correlation) between the activity maps and other images from published articles that directly reference a given term $i$ \cite{yarkoni2011large}. We choose the six terms most clearly associated with resting state activity \textit{default mode}, \textit{salience}, \textit{executive}, these are the resting state network terms and \textit{sensorimotor}, \textit{auditory} and \textit{visual}, sensory network terms. We used these to characterise the mean activity of a given state $s$ by projecting the activity pattern $\mu(s)$ back into 3D brain standard space (see Supplementary Figure \ref{suppfig:neurosynth}A) and inputting the resulting map into \textit{NeuroSynth}. 
	\\*[4pt]
	The resulting score for a state $s$ and term $i$ is denoted $\theta_{i,s}\in[-1,1]$, with 1 indicating perfect correlation between the state's mean functional activity map and $i$ and -1 indicating perfectly anticorrelated activity. We note that although these terms, while chosen to relate to known resting state patterns, are not equivalent and should be thought of as suggestive of a global pattern of activity (or its absence). We explore the activity of actual networks in our spatial community analysis Section \ref{comms}.
	\\*[4pt]
	We propose that the global score $\theta$ can also be considered a dynamically changing property of the system. Given a score $\theta_{i,s}$ for a term $i$ and state $s$, the one step ahead predicted score is
	\begin{align}
	E_{t+1}[\theta_{i,s}] = \sum\limits_{s'\in\mathcal{S}} P_{s,s'} \theta_{i,s'}.\label{eq:pred}
	\end{align} 
	We use this predicted score to examine the global properties of the activity observed after reaching a given state.
	\\*[4pt]
	\textit{NeuroSynth} can also be used in conjunction with the newly developed package \textit{NiMARE} to directly calculate the posterior probability of terms from a large corpus of neuroimaging journal abstracts and images given a selection of brain voxels in standard space \cite{NiMARE53:online}. Due to the variability in brain region size, regions selected by community membership are downsampled by selecting 10,000 voxels with replacement from each community which was found to produce stable posterior probabilities up to the third decimal place.
	\\*[4pt]
	We use \text{NeuroSynth} with \textit{NiMARE} to determine a plausible function for each of our spatial brain region communities, selecting only those terms that are most a posteriori probable and which had a functional rather than anatomical interpretation (see Supplementary Figure \ref{suppfig:neurosynth}B). We pass each community from each hub state through our spatiotemporal community ranking method resulting in a ranked list of communities of brain regions per state and then pass each top ranked community through the \textit{NiMARE/NeuroSynth} method to determine their most likely functional term associations. In order to be comparable with the global score $\theta$, the \textit{NeuroSynth} score is either a positive or negative association depending on the mean activity of the regions as suggested in Section \ref{hubs}.
	\subsection{Validation of Model Framework \label{res:val}}
	A detailed validation of key features of the modelling and analysis framework was carried out using synthetic data (see Supplementary Information, Section \ref{supp:val}). This includes validation of the dimensionality reduction method as a means to reduce the computational demand of modelling while retaining community structure using the Adjusted Rand Index (ARI) \cite{rand1971objective}. Validation is also performed for the Markov Information Graph-based community detection and model selection procedures. Other key components of the model such as the HMM inference procedure have already been validated using synthetic data with detailed simulations \cite{vidaurre2016spectrally,vidaurre2018discovering}.
	\\*[4pt]
	Not all components of the modelling and analysis framework could be validated by simulation as it was considered beyond the scope of this document to generate realistic synthetic community functional homogeneity and \textit{NeuroSynth} scores. The community importance ranking procedure is instead validated using real annotation metadata and the \textit{NeuroSynth} tool.
	\section{Results}
	Results for our multisubject HMM model training and multiplex graph model analysis are given below.
	\subsection{Dimensionality Reduction \label{dimred}}
	We select the appropriate number of principal components using the method of parallel analysis outlined in Supplementary Information, Section \ref{supp:pa}. This resulted in a reduced set of $d=9$ dimensions that account for roughly $75\%$ of the total variance, which are then used in fitting the model. Validation of this approach using synthetic data is explored in Supplementary Information, Sections \ref{supp:pasim} and \ref{supp:cr}. 
	\subsection{Entropy Relates to Model Selection \label{msel_res}}
\begin{figure}[!h]
		\centering
		\includegraphics[width=0.98\textwidth]{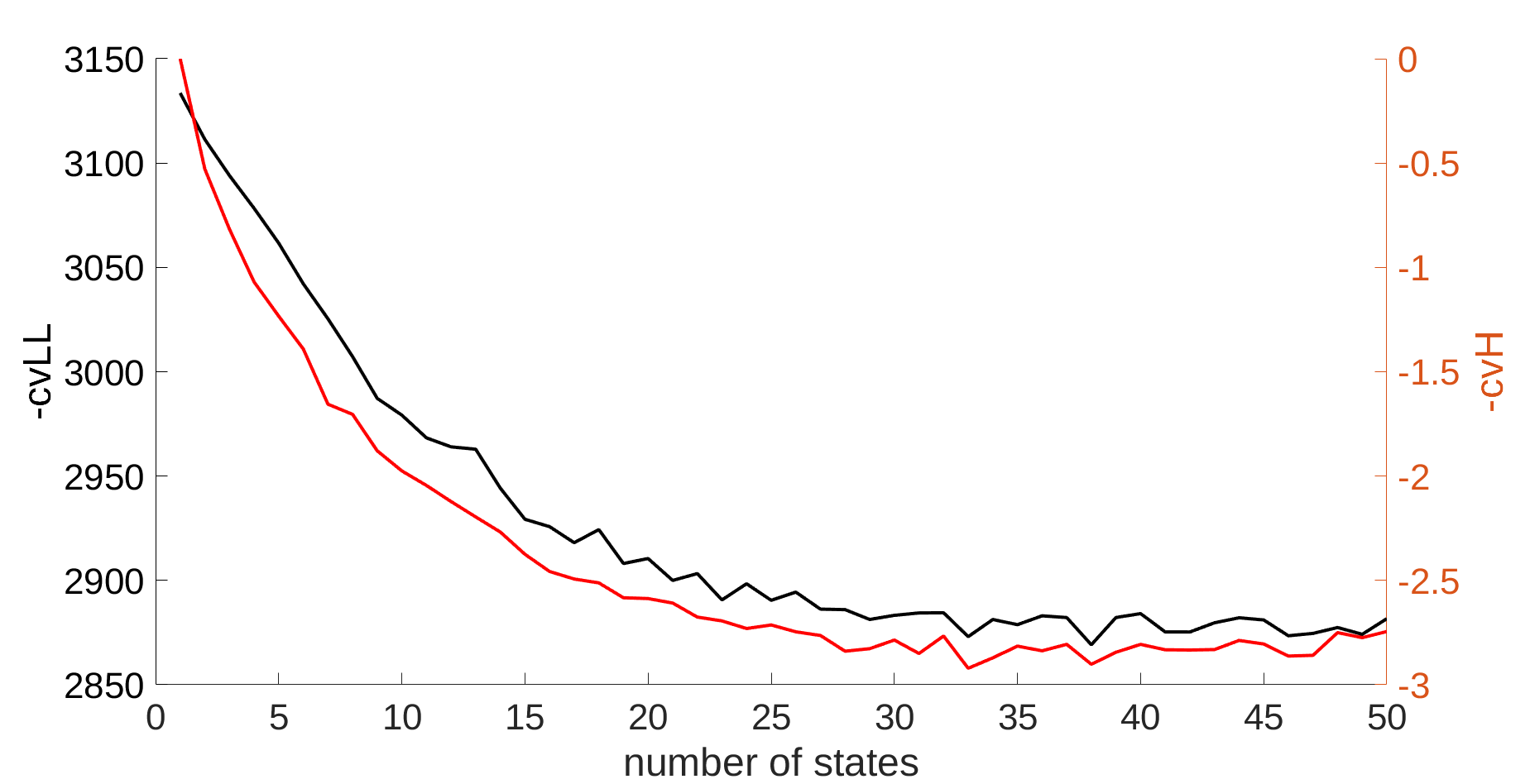}
		\caption{Selection of the number of hidden states by minimising the negative cross-validated entropy. The axes show the negative cross-validated log-likelihood $-\textrm{cvLL}$ (left) and negative cross-validated entropy $-\textrm{cvH}$ (right).\label{suppfig:modsel}. Qualitative similarities are evident between the two criteria suggesting deeper similarities between likelihood and entropy maximisation.}
	\end{figure}

	Applying our cross validated maximum entropy Hidden Markov Model selection criteria by maximising the cross-validated entropy $H(k)$, we obtain an HMM with $K=33$ initial states. Figure \ref{suppfig:modsel} shows that the entropy maximum also coincides with the maximisation of the cross-validated Bayesian log-likelihood, which is a general indicator of model fit. To further reduce the risk of overfitting, we exclude those states that occur in less than 25$\%$ of subjects and renormalise $P$ so that the rows again sum to one. The resulting model has a total of $K=27$ brain states.
	\subsection{Network Dynamics Indicate Clustering of Activity Patterns in Space and Time}
	Table \ref{tab:neurofor} shows that states positively correlated with resting state activity terms are significantly more likely to transition to states with similar associations and vice versa (see Supplementary Figure \ref{suppfig:neurofor} for linear model comparison). In contrast, states correlated with resting state terms tended to transition to states that are negatively correlated with the sensory terms. This suggests that states associated with the former resting state networks tend to co-occur to the exclusion of sensory and sensorimotor patterns of activity. These results indicate a spatiotemporal separation between resting state network activity and sensory activity.
	\\*[4pt]
	States with high scores for sensory activity terms show a far weaker positive affinity for transition to each other than do the former resting state network terms. This suggests that concurrent activity in space and time is most likely between states with high resting state network activity. This pattern of concurrent activity is only weakly suggestive for sensory modes of activity. In contrast, robust mutually antagonistic spatiotemporal relationships between sensory and resting state network associations are present. We shall see in Section \ref{comms} this pattern of mutual exclusivity is mirrored by the most central states in the network or hub states at both the global (functional activity map) and the local (network community) levels. States show a general trend of transitioning from terms with one global activity association to another state that scores highly for the same association, suggesting some level of brain state inertia in the global pattern of functional activity. 
	\begin{table}[!h]
		\centering
		\begin{tabular}{lllllll}
			\textbf{}  & DM          & S         & E         & SM        & V           & A         \\\hline
			DM$_{t+1}$ &  \textcolor{red}{0.9866} **  & \textcolor{red}{0.8201} ** & \textcolor{red}{0.4601} *  & \textcolor{blue}{-0.4416} * & \textcolor{blue}{-0.3939} (*) & \textcolor{blue}{-0.5488} ** \\
			S$_{t+1}$ & \textcolor{red}{0.8253} **  &   \textcolor{red}{0.9863} **         & \textcolor{red}{0.3550} (*) & \textcolor{blue}{-0.5526} ** & \textcolor{blue}{-0.5302} *   & \textcolor{blue}{-0.481} *  \\
			E$_{t+1}$  & \textcolor{red}{0.4519} *    & \textcolor{red}{0.3615} (*) &     \textcolor{red}{0.9826} **      & \textcolor{blue}{-0.5203} * & \textcolor{blue}{-0.5241} *   & \textcolor{blue}{-0.5753} ** \\
			SM$_{t+1}$ & \textcolor{blue}{-0.4576} *   & \textcolor{blue}{-0.5781} **  & \textcolor{blue}{-0.5215} * &   \textcolor{red}{0.9878} **        & \textcolor{red}{0.1765}      & \textcolor{red}{0.3128}    \\
			V$_{t+1}$  & \textcolor{blue}{-0.3718} (*) & \textcolor{blue}{-0.5085} *  & \textcolor{blue}{-0.5092} * & \textcolor{red}{0.1683}    &  \textcolor{red}{0.9885} **           & \textcolor{red}{0.1773}    \\
			A$_{t+1}$  & \textcolor{blue}{-0.4874} *   & \textcolor{blue}{-0.4226} *  & \textcolor{blue}{-0.5364} * & \textcolor{red}{0.2657}    & \textcolor{red}{0.1733}      &  \textcolor{red}{0.9842} **        
		\end{tabular}
		\caption{This table shows the relationships between \textit{NeuroSynth} terms scores, calculated using the mean activity brain map for each state and the one step ahead projected score for each term according to the model (see Equation \ref{eq:pred}). Term scores for each state are correlated with the projected term scores one time step into the future (denoted by subscript $t+1$) from the current state (red is positive correlation, blue negative). False discovery rate corrected t-test significance is marked as ** ($p<0.01$), * ($p<0.05$) and (*) for marginal results ($p<0.1$).  The comparison between state scores for each term and the one step ahead predicted scores shows that there is a spatiotemporal relationship between resting state terms which are anticorrelated with sensory terms. \label{tab:neurofor}}
	\end{table}
	\subsection{Evidence for Metatastate Structure in Wakefulness}
\begin{figure}[!h]
		\centering
		\includegraphics[width=0.98\textwidth]{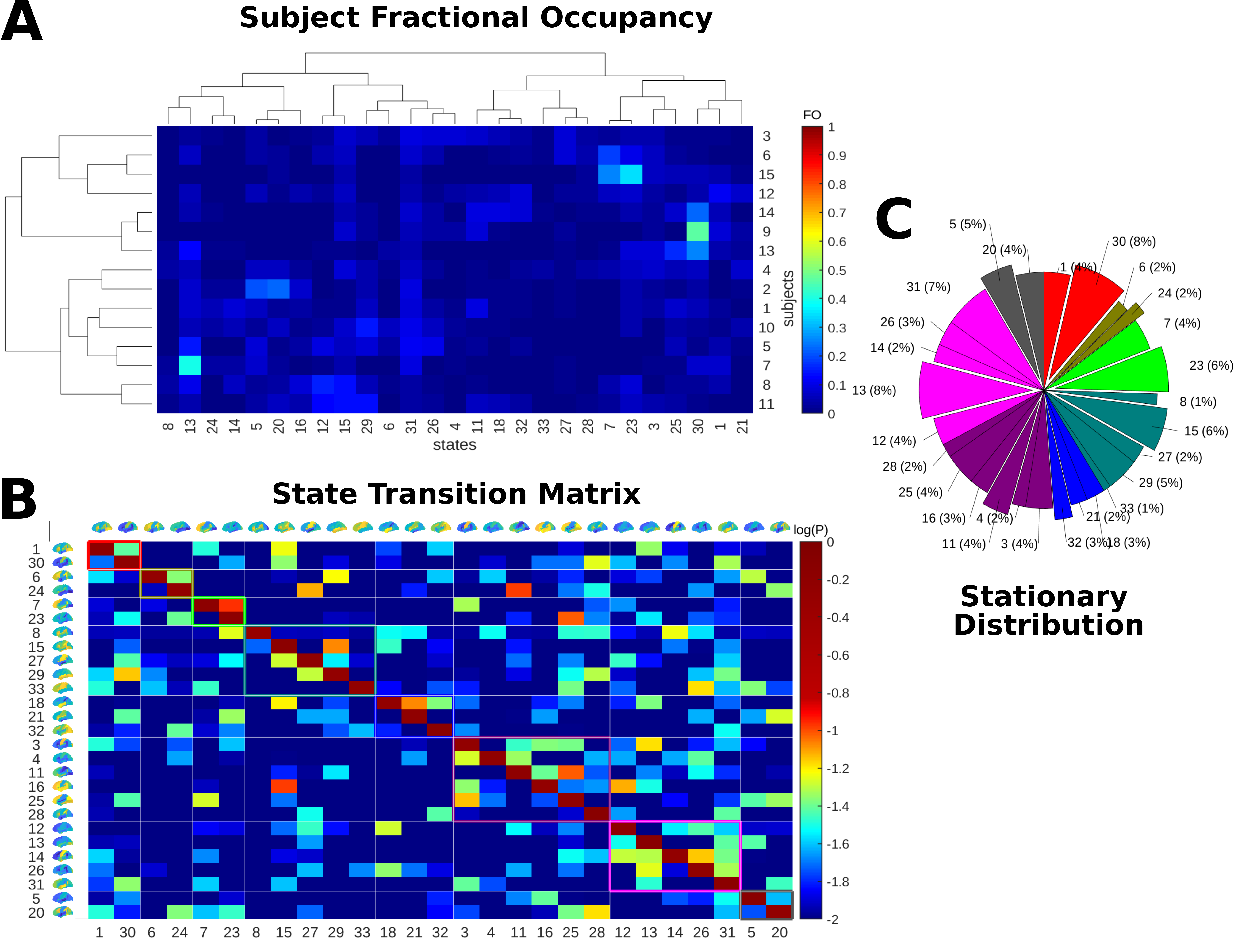}
		\caption{A summary of the subject state network dynamics. \textbf{(A)} Clustergram showing the relationships in Fractional Occupancy (FO), the proportion of time spent in a state clustered by subjects (vertically) and by states (horizontally). \textbf{(B)} The log of the state transition matrix $P$ is shown, where states have been grouped along the diagonal, according to their community membership. \textbf{(C)} Pie chart showing the state occupancy at equilibrium (the probability of finding a subject in a state in the limit as time goes to infinity). Wedges in this pie chart are the individual hub states in each community according to the community $z$-score. These two scores share a significant 0.537 ($p=0.004$), indicating the importance of community centrality to long run behaviour.  \label{fig:wakesumm}}
	\end{figure}

	In order to demonstrate the presence of temporal community structure, we performed hierarchical linkage clustering using the correlation in $\kappa$ between subjects and states. We also calculated the normalised degree of symmetry in $P$, $\textrm{Sym}(P)=0.9921$ indicating a degree of symmetry in $P$ (with $\textrm{Sym}(P)=1$ when $P$ is completely symmetric). Figure \ref{fig:wakesumm}A suggests a temporally clustered pattern of state fractional occupancy in which certain states are more likely to co-occur in one subset of subjects than in the other. Figure \ref{fig:wakesumm}B shows the transition probability matrix $P$ organised into communities by Louvain community detection, where $\gamma=0.48$ (as selected by Variation of Information minimisation) \cite{lambiotte2008laplacian}. Temporal communities indicate modules of clustered state transitions. This temporal community partition was tested for robustness by comparing the $Q$ modularity statistic to 10,000 random partitions with the same community labels ($p=$1e-4). 
	\\*[4pt]
	Each community, $U\subset \mathcal{S}$, is characterised by a hub state $h(U)$ determined by the state with the highest community degree $z$-score, a measure of state centrality to the temporal network (see Supplementary Figure \ref{suppfig:hubs}). Figure \ref{fig:wakesumm}C, shows the long run probability of state $s$ re-occurence $\pi_s$. Re-occurence and centrality to a community appear to be strongly correlated as states more central to their communities according to the $z$-score, $z(s)$, also tended to have a higher stationary probability $\pi_s$, with correlation coefficient $\rho=0.537$ ($p=0.004$). This observation suggests that as mediators of network dynamics, community hub states tend to re-occur, playing a central role in the overall network dynamics as well as in their own community.
	\subsection{Community Rankings Reveal Spatiotemporal Modules of Functional Activity \label{comms}}
	Louvain community detection was performed for each of the community hub state graphs $G(h(U))$ for each community $U$ in partition $\mathcal{U}$. We assessed the degree of symmetry in the Markov Information graph of each hub states and found that $\textit{Sym}(W(h(U)))>0.99$ for all communities $U$. Here, the Variation of Information was not used to select $\gamma$ as differing recommended $\gamma$ between hubs was found to produce communities of inconsistent and incomparable sizes; we thus select the resolution as $\gamma=2$ for all hub states. This was found to produce median spatial community network sizes that were sufficiently small on average (roughly 4 regions per community) for our community ranking method to efficiently sample the graph while also being large enough to detect functionally conserved brain state networks.
	\\*[4pt]
	We perform \textit{NeuroSynth} analysis by taking the mean functional activity maps generated for the hub states as input in combination with the resting state network terms \textit{default mode}, \textit{salience}, \textit{executive} and the sensory network terms \textit{sensorimotor}, \textit{auditory} and \textit{visual} (see Supplementary Figure \ref{suppfig:neurosynth}A for algorithmic explanation).  The results in Table \ref{tab:Neurosynth} suggests a separation between sensory and resting state activity in space and time with hub states scoring highly for either resting state or sensory terms but rarely both. Table \ref{tab:Neurosynth} gives the highest ranked functional terms (filtering out purely anatomical terms) in each hub state for the top three ranked spatial network communities (using our ranking method). The top terms for each of the networks (communities) in the states largely coincide with the functional associations ascribed to each of the hub states themselves.
	\\*[4pt]
	Exploring these relationships, we see that in some cases the connections between spatial community function and hubs are direct. State 23 shows a positive association with \textit{observation} and \textit{action} in dominant spatial communities and a strong association with all three sensory network terms. State 11 shows a clear association with auditory activity as well as a top ranked community association with the term \textit{voice}. In state 15, which shows a strong correlation with $\textit{visual}$ activity, the top ranked communities include positive associations with the \textit{face} (a common object of visual processing). 
	\\*[4pt]
	In some states we see both strong positive and negative associations. Global negative asssociations are difficult to interpret in isolation as evidence of anticorrelated network behaviour within a state, however when paired with mesoscale information from the top ranked communities a stronger case is possible. State 32 appears mixed in activity but shows strong to moderate negative correlations with visual and auditory processing. The latter of these is corroborated by the anticorrelated speech network. State 30 is another state with mixed associations based purely on global functional activity, however, we see both moderate negative correlation globally with visual activity, and a specific negatively correlated community related to visual tasks or processing, suggesting a visual down state. A similar explanation can be used for state 5. State 23 is a sensory associated state with sensory associations at both the global and network scales. State 23 is negatively correlated with default mode activity. The default mode network is involved in language comprehension and reasoning, explaining the anticorrelated network associated with syntactic processing. Negatively associated communities may more generally suggest decreased metabolic or functional demand for these in networks leading to a coordinated down state.
	\begin{table}[!h]
		\resizebox{\textwidth}{!}{
			\begin{tabular}{lcccccccc}
				\textbf{term} &                &         &     &  \textbf{hub state}            &             &         &                  &         \\\hline
				& 13             & 11      & 15        & 32               & 23          & 5       & 24               & 30      \\
				DM   & -0.0707        & 0.0528  & -0.0094   & 0.1085           & -0.3496     & -0.0341 & -0.095           & -0.1918 \\
				S       & -0.0924        & 0.0506  & -0.1099   & 0.0714           & -0.2039     & 0.0164  & 0.026            & -0.0942 \\
				E      & -0.2946        & -0.1984 & 0.0534    & 0.1736           & -0.085      & 0.1017  & -0.0772          & 0.1024  \\
				SM   & 0.4245         & -0.1793 & -0.0535   & 0.1239           & 0.2502      & -0.3186 & 0.2973           & 0.179   \\
				V         & 0.1544         & -0.0055 & 0.4604    & -0.536           & 0.2308      & -0.289  & -0.1644          & -0.1168 \\
				A       & 0.1574         & 0.2959  & -0.0946   & -0.1542          & 0.2725      & 0.039   & 0.1534           & 0.0306  \\
				\textbf{rank} &          &    & & \textbf{community} & & &  &   \\\hline
				1 & -reward         & +voice   & -incentive & -speaker          & +action      & -visual  & -autobiographical & -basal   \\
				2 & -theory of mind & +memory  & +action    & +autobiographical & +observation & +memory  & -empathic         & -memory  \\
				3 & -language       & -action  & +face      & +syntactic        & -syntactic   & +voice   & -autonomic        & -visual 
			\end{tabular}
		}
		\caption{Summary of the \textit{NeuroSynth} results for the hub states. The terms scored by \textit{NeuroSynth} are the resting state terms \textit{default mode} (DM), \textit{salience} (S), \textit{executive} (E) and sensory terms \textit{sensorimotor} (SM), \textit{visual} (V) and \textit{auditory} (A). The first rows of the table under hub states show the \textit{NeuroSynth} correlation score between each of the hub states' brain maps and the terms on the left (see Supplementary Figure \ref{suppfig:neurosynth}). The second section under terms shows the most probable terms associated with each of the top three communities identified by our ranking method, providing further information on the component functional communities of these states. The sign next to each term indicates whether the association is positive or negative (depending on the sign of the brain regions involved).\label{tab:Neurosynth}}
	\end{table}
	\section{Discussion \label{methdisc}}
	In this paper we present a fully unsupervised pipeline for characterising the spatiotemporal activity of neuronal brain states in terms of a multiplex brain state graph model. This pipeline involves the training of an HMM in order to obtain a multiplex spatiotemporal directed brain state graph that represents the dynamics of subjects in resting wakefulness. We present a method for obtaining a set of states (layers) that generalises well over subjects and use this method to determine key states in the network dynamics. Lastly, we characterise the spatiotemporal components of the model that are most central and most functionally coherent, characterising these using metatextual image analysis of the neuroscience literature.
	\\*[4pt]
	Our HMGM-based methodology reveals a rich array of complementary communities acting together to produce modes of neural behaviour during resting wakefulness. Crucially, we have shown that patterns of activity resembling the resting state networks tend to co-occur and that these patterns tend to preclude sensory and sensorimotor patterns of activity. This modularity of brain state function has been suggested by others \cite{smallwood2012cooperation,vidaurre2017brain}, but metaanalysis of terms associated with these functions allows us to characterise individual states and quantify their change in character through time.
	\\*[4pt]
	Within each hub brain state the division between functions was not clearly partitioned, with many terms featuring communities with \textit{memory} or \textit{autobiographical} associations, possibly suggesting an undercurrent of narrative thought which persists across numerous states. Alternatively, this may be due to artefacts caused by auditory memory-related tasks studies in the \textit{NeuroSynth} database. It is important to note that spatiotemporal state-based activity analysis is novel and so terms in the literature which derive from static models of activity may not map accurately onto dynamic patterns of activity. In particular, transient states may be smoothed out of these analyses meaning that new studies will need to be performed focusing on dynamic functional activity change at much shorter time scales in order to build up an understanding of function in dynamic brain states.
	\\*[4pt]
	Some of the state global functional activity term associations, particularly negative ones, remain difficult to interpret. In state 13, there is a strong association with the term \textit{sensorimotor}, however all of the top ranked communities for this state are negatively associated with functions that may have a closer association to resting state activity. This could be due to putative link between the central executive activity and reward observed in primates \cite{sigmund2001reward}, but may also be due to ranking error or noise in our graph model. However, the roles of many states become more clear when combining functional information from either anticorrelated or correlated mesoscale communities with global tendencies in functional activity. We hypothesize that strongly cohesive anticorrelated networks may be entering a coordinated down state due to changes in metabolic or functional demand \cite{tomasi2017dynamic,passow2015default,thompson2018neural}.
	\\*[4pt]
	One issue with our approach is that the Louvain implementation we use with directed modularity does not fully capture the signal of edge directionality in community detection (see Supplementary Information Section \ref{supp:mod}). This problem may be partially mitigated by the fact that we found the edge weights in question to not be highly asymmetric when measured as a fraction of matrix energy. However, a community detection methods that more directly account for directed edges, such as InfoMap \cite{rosvall2009map}, or the Markov structure of the model, such as \cite{jin2011ant} may identify other other forms of community structure in our graph models that are worth investigation. In particular we intend to investigate more general implementations of the Louvain algorithm that are optimised for directed networks \cite{li2018improved,dugue2015directed}.
	\\*[4pt]
	Presently, our framework also does not fully take advantage of the multuiplex graph structure of the model, for example using multilayer community detection which can be complex to parametrise \cite{hanteer2020unspoken}. However, a potential advantage of the HMGM framework is that it provides a way to ground the interlayer coupling parameters used in some multilayer community detection using a natural property of the model, the probability of state transition. In our future work we intend to investigate multilayer community detection approaches to look at dynamic changes in network membership using coupling parameters based on the transition probabilities between state layers.  
	\\*[4pt]
	We plan to apply our multiplex analysis framework to conditions of altered consciousness in deep anaesthesia and determine novel spatiotemporal networks that characterise this condition with comparison to our current graph model for resting wakefulness. In this way we hope to elucidate the complex network dynamics underlying conscious brain activity \cite{huang2021survey}.
	
	\begin{backmatter}
		
		%
		%
		\section*{Abbreviations}
		\textbf{BOLD:} Blood Oxygen Level Dependent signal\\
		\textbf{CEN:} Central Executive Network\\
		\textbf{DMN:} Default Mode Network\\
		\textbf{fMRI:} functional MRI (Magnetic Resonance Imaging)\\
		\textbf{FO:} Fractional Occupancy\\
		\textbf{PCA:} Principal Component Analysis\\
		\textbf{HMM:} Hidden Markov Model\\
		\textbf{ROI:} Region of Interest\\
		\textbf{SM:} Sensorimotor\\
		\textbf{SN:} Salience Network
		\section*{Availability of data and materials}
		Data, community ranking, and model selection code is available from the authors upon request.
		\section*{Ethics approval and consent to participate}
		The study was approved by the Local Research Ethics Committee (Oxford Research Ethics Committee B, Oxford, UK) and data collection was performed between October and December 2009. The study was performed in line with the Declaration of Helsinki and all subjects gave written informed consent.
		\section*{Competing interests}
		The authors declare that they have no competing interests.
		%
		%
		\section*{Funding}
		JBW is supported by the Commonwealth Scholarship Commission UK, the Ernest Oppenheimer Memorial Trust (South Africa) and Human Brain Project, Specific Grant Agreement 3 (award reference 945539), CEW is funded by  MRC Development Pathway Funding Scheme (award reference MR/R006423/1), and GDR is partially supported by the UK Engineering and Physical Sciences Research Council (EPSRC) grants EP/R018472/1 and EP/T018445/1. This research is funded in part by the Wellcome Trust (grant 203139/Z/16/Z). For the purpose of open access, the authors have applied a CC-BY public copyright license to any Author Accepted Manuscript version arising from this submission. Data collection was funded by the National Institute for Academic Anaesthesia, and the International Anaesthesia Research Society. 
		\section*{Authors' contributions}
		JW prepared the draft manuscript and developed the analysis methods. CW, CD and GR edited the manuscript. CW provided the raw data and interpretation of neuroscientific results. CD and GW supervised the analytical methods development. GR contributed to interpretation of model results. All authors read and approved the final manuscript.
		\section*{Acknowledgements}
		We would like to thank the reviewers and editors for their helpful suggestions in restructuring and correcting this manuscript. We are grateful to the attendees and organisers of the Communities in Networks conference, where this work was originally presented, for the opportunity to contribute to this Special Issue. We are also grateful to Mark Woolrich, Angus Stevner, and the members of the Oxford Anaesthesia Neuroimaging and Protein Informatics Groups, for their insightful questions and comments. 
		
		%
		
		
		\bibliographystyle{bmc-mathphys} 
		\bibliography{bmc_article}      
		
		
		
		
		%
		%
		%
		
		
		\section*{Additional Files}
		\subsection*{Additional file 1 --- Supplementary Information}
		Supplementary Methods, Figures and Tables.
		
	\end{backmatter}
\clearpage
\setcounter{equation}{0}
\setcounter{figure}{0}
\setcounter{table}{0}
\makeatletter
\renewcommand{\theequation}{S\arabic{equation}}
\renewcommand{\thefigure}{S\arabic{figure}}
\section*{Supplementary Information}
\section{Parallel Analysis for Dimensionality Reduction \label{supp:pa}}
Let $X$ be a $F\times T$ multivariate matrix with $F$ features and $T$ time points. Dimensionality reduction of $X$ by Principal Component Analysis (PCA) requires the selection of the reduced dimension $d<D$. Parallel analysis allows this to be done in a data driven way by comparing the original data set to surrogate data \cite{horn1965rationale}. In parallel analysis, PCA is first performed on $X$. The resulting eigenvalues can be ordered $\lambda_{(d)}$ so that $d$ is the $d$ largest eigenvalue. 
\\*[4pt]
Next, the columns (time points) of $X$ are permuted within each row removing structure from the dataset and this process is repeated $R=10000$ times producing $X_1,..,X_R$ surrogate data sets with the same row-wise distribution as $X$. For each $X_r$ we can obtain a corresponding $d$ largest eigenvalue $\hat \lambda_{(d),r}$. 
\\*[4pt]
The optimum choice for $d$ is given by the smallest $d$ satisfying
\begin{align*}
\lambda_{(d+1)}<\hat\lambda_{(d+1)}^{P^{th}},
\end{align*}
where $\hat \lambda_{(d)}^{P^{th}}$ is the $P^{th}$ of the permuted eigenvalues $\hat \lambda_{(d),1},...,\hat \lambda_{(d),R}$. The value of $P$ determines how much the eigenvalues of the components of $X$ must dominate the eigenvalues of the permuted datasets. We choose the percentile $P=99$. In other words, the first $d$ components in the original dataset must each account for more variance than 99\% of the permuted components. This was chosen rather than the standard $P=95$ in order include as much of the signal in $X$ as possible for HMM model training. The dimensionally reduced data set $X^*$ is thus given by
\begin{align*}
X^*=AX
\end{align*}
where $A$ is the eigenmatrix of the first $d$ columnwise eigenvectors of $X$.
\section{Directed, Weighted Modularity Score \label{supp:mod}}
The directed modularity score $Q(\mathcal{C})$ for a given partition $\mathcal{C}\subset 2^{V}$ of a weighted, directed graph $G=(V,A)$ with node set $V$ and adjacency matrix $A$ is a measure of how well the partition separates nodes into modules by highly scoring partitions with lower weights on between community edges and higher weights on within community edges. It is calculated as
\begin{align}
Q(\mathcal{C}) &= \frac{1}{m}\sum\limits_{v,v'\in V}\left[A_{v,v'}-\gamma \frac{k_v^{out}k_{v'}^{in}}{m}\right]\delta_\mathcal{C}(v,v')\\
&=\frac{1}{m}\sum\limits_{v,v'\in V}B_{v,v'}\delta_\mathcal{C}(v,v'),\label{supp:Q(C)}
\end{align}
for $m=\sum\limits_{w,w'\in V} A_{w,w'}$, $k^{in}_{v'}=\sum_{u\in V}A_{u,v'}$, $k^{out}_{v}=\sum_{u\in V}A_{v,u}$ and $\delta_\mathcal{C}(v,v')$ is the Dirac delta function that is one if and only if their exists $C\in \mathcal{C}$ such that $v,v'\in C$ and zero otherwise \cite{newman2006modularity}. The matrix $B$ with elements
\begin{align*}
B_{v,v'}=A_{v,v'}-\gamma \frac{k_v^{out}k_{v'}^{in}}{m},
\end{align*}
is known as the directed modularity matrix of $A$ \cite{nicosia2009extending,leicht2008community}.
\\*[4pt]
The Louvain implementation we use requires that the modularity matrix $B$ be symmetric. This is achieved by symmetrising, $B'=(B+B^T)/2$ so that
\begin{align*}
Q'(\mathcal{C}) &=\frac{1}{m}\sum\limits_{v,v'\in V}B'_{v,v'}\delta_\mathcal{C}(v,v')\\
&=\frac{1}{2m}\sum\limits_{v,v'\in V}\left[A_{v,v'}-\gamma \frac{k_v^{out}k_{v'}^{in}}{m}+A_{v',v}-\gamma \frac{k_{v'}^{out}k_v^{in}}{m}\right]\delta_\mathcal{C}(v,v'),\\
&=Q(\mathcal{C}),
\end{align*}
resulting in the directed modularity score as presented in Equation \eqref{supp:Q(C)}.
\section{Measuring the Symmetry of Markov Matrices \label{supp:asym}}
It is useful to have a measure to assess the 'degree of symmetry' in a directed network. One way to do this is to consider the energy (as measured by the Frobenius norm) in the weight matrix $W$ which is symmetric \cite{aggarwal2020linear}.
\begin{align*}
\textrm{Sym}(W)=\frac{1}{4}\frac{||W+W^T||_F^2}{||W||_F^2},
\end{align*}
where $||\cdot||_F$ is the Frobenius norm and $||W+W^T||_F$ is the symmetric part of $W$. The measure $0\le \textrm{Sym}(W)\le 1$ is near to $0$ when $W$ is 'highly asymmetric' (minimised when $W$ is skew-symmetric i.e. $-W=W^T$) and near to $1$ when $W$ is 'highly symmetric' (maximised when $W=W^T$). 
\section{Community Centrality \label{supp:centrality}}
The community centrality for the temporal graph $G(P)=(\mathcal{S},P)$ is calculated using the symmetric undirected version of the transition matrix $P$ to obtain the within community degree centrality z-score, $z(s)$ for $s\in\mathcal{S}$ \cite{guimera2005functional}. Given a partition $\mathcal{U}\subset 2^{\mathcal{S}}$ of the temporal graph into communities, $U$, this statistic measures how well connected $s\in U$ is in relation to the rest of $U$. The undirected network is based on $G(P')=(\mathcal{S},P')$, where $P'=(P+P^T)/2$. The score is
\begin{align*}
z(s) = \frac{\nu_s-\nu_U}{\tau_U},
\end{align*}
where $\nu_s$ is the community-specific degree
\begin{align*}
\nu_s = \sum\limits_{s'\in U-s} P'_{s,s'},
\end{align*}
and $\nu_U$ is the the expected centrality over all other nodes in $U$,
\begin{align*}
\nu_U = \frac{1}{|U|}\sum\limits_{s\in U} \nu_s.
\end{align*}
Lastly $\tau_U$ is the standard deviation of $\nu_s$ for $s\in U$, so
\begin{align*}
\tau_U = \sqrt{\frac{1}{|U|}\sum\limits_{s,s'\in U} (\nu_s-\nu_U)^2}.
\end{align*}
\begin{figure}[h!]
	\includegraphics[width=\textwidth]{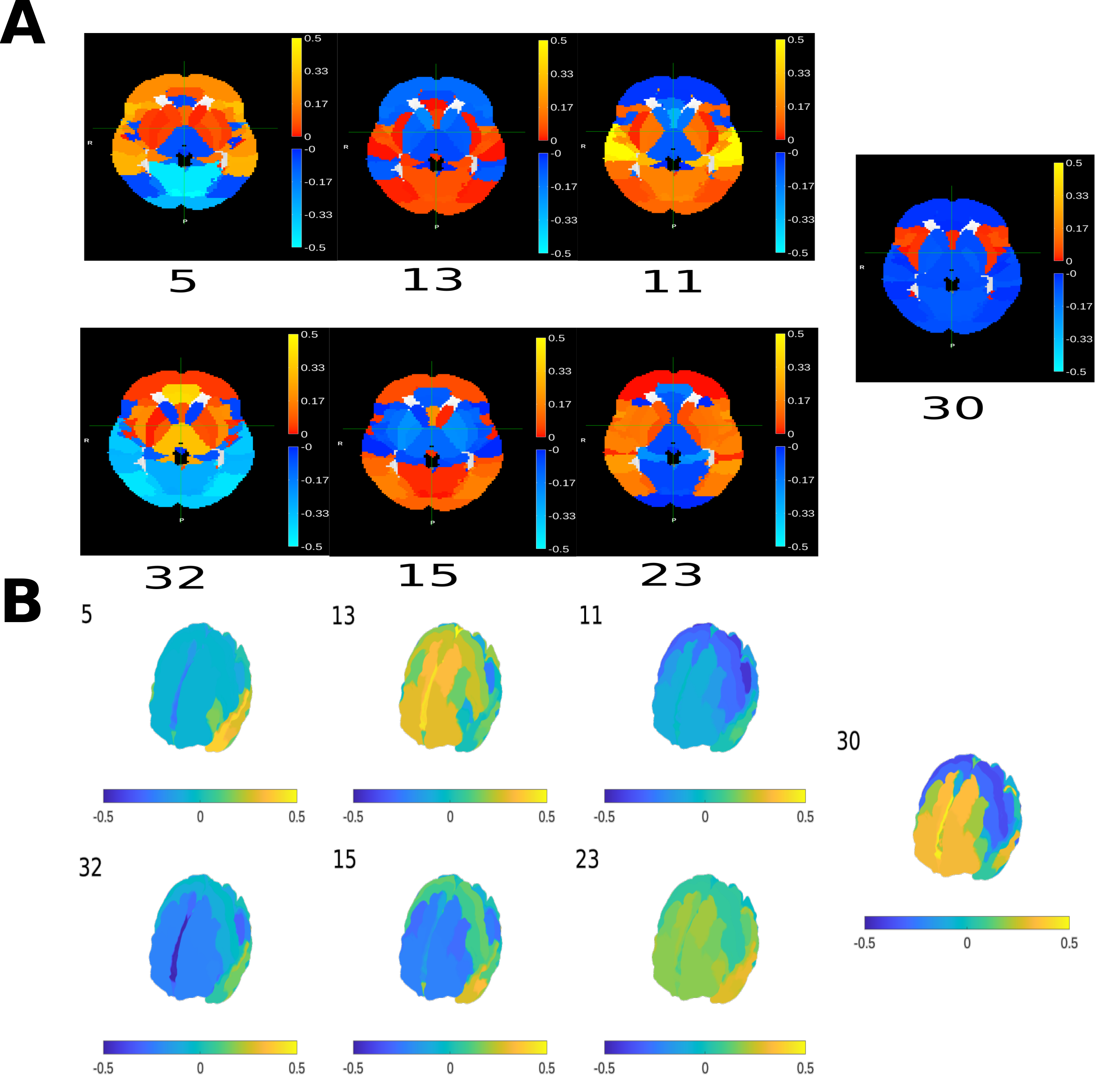}
	\caption{Hub state mean activity brain maps. (A) The figure shows activity for a central axial slice of the 3D mean activity brain maps of all hub states (the index is determined by the original 33 state HMM). (B) Plots of a surface representation of the same mean activity brain maps for the same hub states. \label{suppfig:hubs}}
\end{figure}
\begin{figure}[!h]
	\centering
	\resizebox{\textwidth}{!}{
\begingroup%
  \makeatletter%
  \providecommand\color[2][]{%
    \errmessage{(Inkscape) Color is used for the text in Inkscape, but the package 'color.sty' is not loaded}%
    \renewcommand\color[2][]{}%
  }%
  \providecommand\transparent[1]{%
    \errmessage{(Inkscape) Transparency is used (non-zero) for the text in Inkscape, but the package 'transparent.sty' is not loaded}%
    \renewcommand\transparent[1]{}%
  }%
  \providecommand\rotatebox[2]{#2}%
  \newcommand*\fsize{\dimexpr\f@size pt\relax}%
  \newcommand*\lineheight[1]{\fontsize{\fsize}{#1\fsize}\selectfont}%
  \ifx\svgwidth\undefined%
    \setlength{\unitlength}{1380.00002884bp}%
    \ifx\svgscale\undefined%
      \relax%
    \else%
      \setlength{\unitlength}{\unitlength * \real{\svgscale}}%
    \fi%
  \else%
    \setlength{\unitlength}{\svgwidth}%
  \fi%
  \global\let\svgwidth\undefined%
  \global\let\svgscale\undefined%
  \makeatother%
  \begin{picture}(1,0.5451081)%
    \lineheight{1}%
    \setlength\tabcolsep{0pt}%
    \put(0.15777573,0.4060025){\makebox(0,0)[lt]{\smash{\begin{tabular}[t]{l}\huge $Pr(T|D)=\frac{Pr(D|T)Pr(T)}{Pr(D)}$\end{tabular}}}}%
    \put(0,0){\includegraphics[width=\unitlength,page=1]{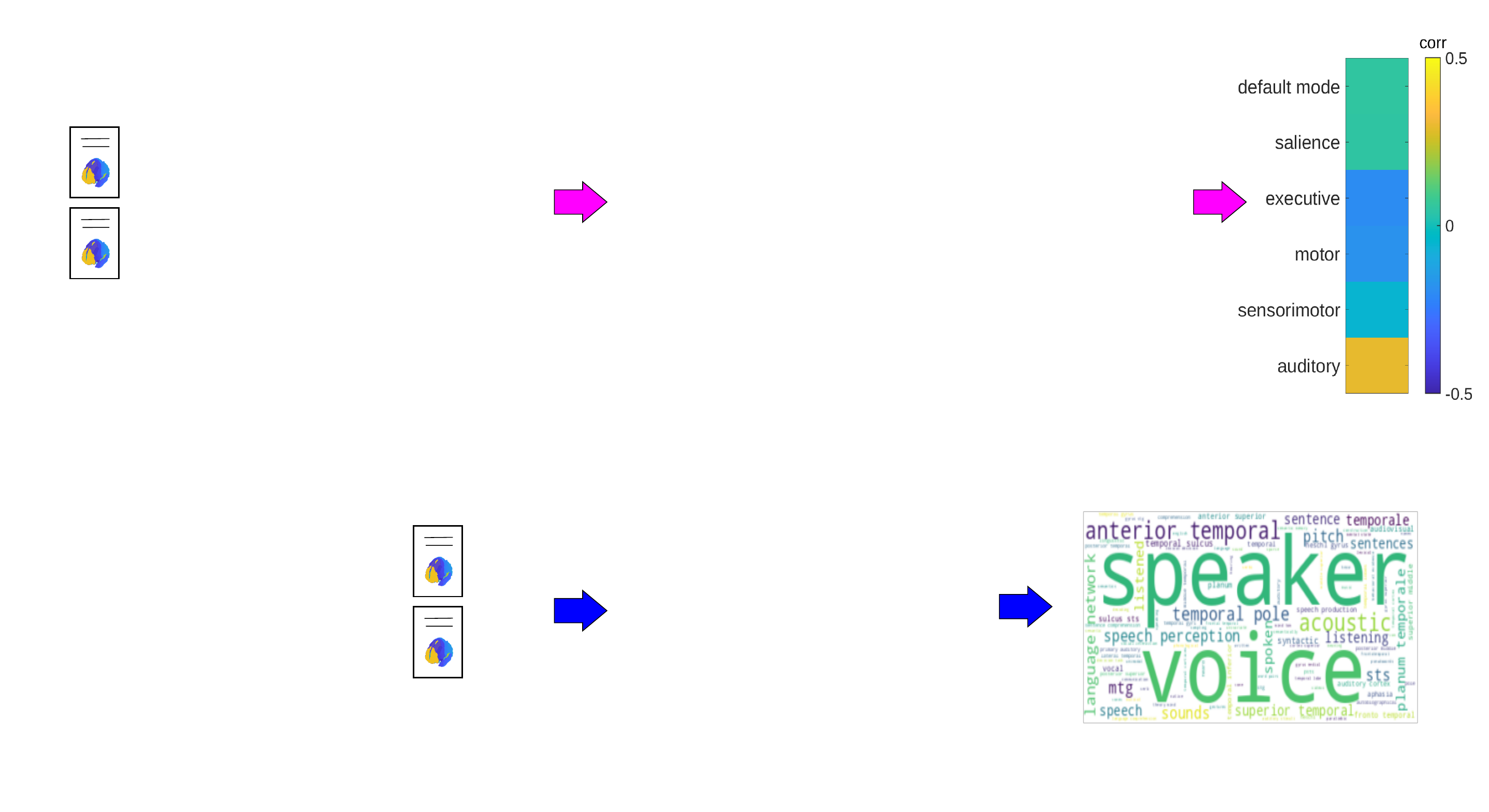}}%
    \put(0.420144,0.13383373){\makebox(0,0)[lt]{\smash{\begin{tabular}[t]{l}\huge $Pr(T|C,D)=\frac{Pr(T|C,D)Pr(T)}{Pr(C,D)}$\end{tabular}}}}%
    \put(0,0){\includegraphics[width=\unitlength,page=2]{neurosynth.pdf}}%
    \put(0.12472694,0.23853919){\makebox(0,0)[lt]{\smash{\begin{tabular}[t]{l}\Huge \textit{Neurosynth}+NIMARE\textbf{(Community,Data) $\rightarrow$ $Pr(\textrm{Term}|\textrm{Community, Data})$ }\end{tabular}}}}%
    \put(0.1163697,0.52859151){\makebox(0,0)[lt]{\smash{\begin{tabular}[t]{l}\Huge \textit{Neurosynth}\textbf{(Term,State$|$Data) $\rightarrow$ Corr(Term,State)}\end{tabular}}}}%
    \put(0,0){\includegraphics[width=\unitlength,page=3]{neurosynth.pdf}}%
    \put(0.10480708,0.05477884){\makebox(0,0)[lt]{\smash{\begin{tabular}[t]{l}\huge Temporal\\\huge Lobe\end{tabular}}}}%
    \put(0,0){\includegraphics[width=\unitlength,page=4]{neurosynth.pdf}}%
  \end{picture}%
\endgroup%
}
	\caption{Diagram overviewing the two \textit{Neurosynth} methods used in our analysis, using the specific example of State 11, Community 1. (A) The first method shows how \textit{Neurosynth} (NS) can be used to generate correlation scores from brain state activity maps and a set of predefined brain terms. In order to perform such a correlation analysis \textit{Neurosynth} requires a corpus of data ($D$) composed of abstracts with associated brain activation coordinates in $(x,y,z)$ voxel space. We used data from Version 7, dated July 2018 which includes 14,371 studies with 3,178 terms drawn from the data (after removal of numeric characters and words with a frequency less than 1 in 1,000). From this corpus a posterior distribution of term ($T$) association at each voxel is produced through a Naive Bayes scheme. Term activity maps for a prespecified term are then correlated with the state mean activity brain map and the whole process is repeated for each of the terms of interest. Finally, a full profile of correlations for each term is outputted. (B) For this method we employ the \textit{NiMARE} tool (NM) which uses the \textit{Neurosynth} algorithm to produces a posterior probability over all terms in the corpus for a given selection of voxels. We use as input the voxels defined by the spatial community of brain regions outputted by the community ranking procedure (spatial Community 1 of State 11). The spatial community ($C$) has been binarised and projected onto the brain map. We show as reference that community activity seems mostly to be located in the temporal lobe, a key region for auditory processing. The final posterior probability of term association is shown as a word cloud where the size of terms is proportional to their predicted probability of association. This method is available in the \textit{NiMARE} Python package. \label{suppfig:neurosynth}}
\end{figure}
\clearpage
\begin{sidewaysfigure}[!h]
	\centering
	\includegraphics[width=\textwidth]{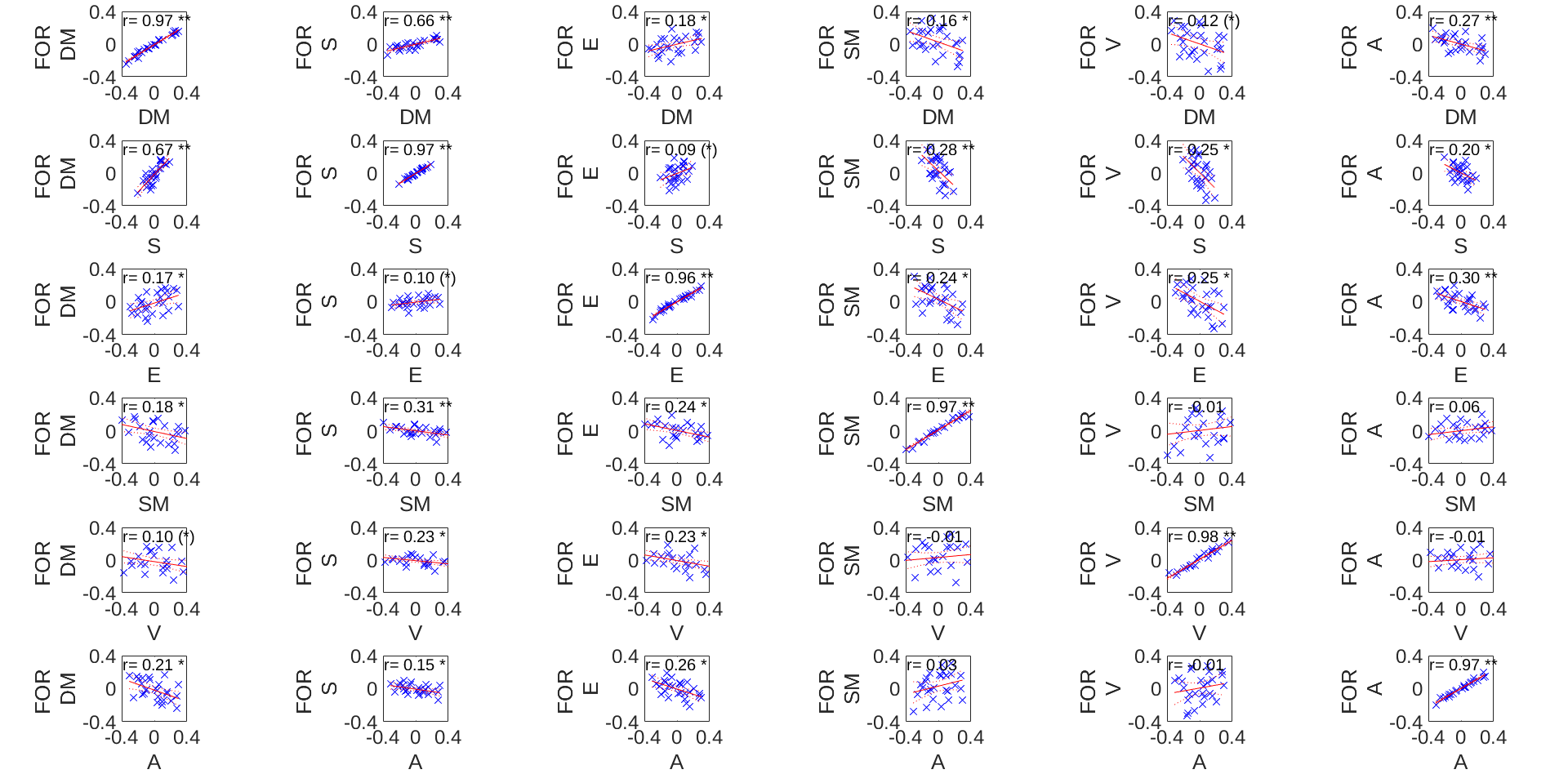}
	\caption{Grid of scatter plots showing the relationships between \textit{Neurosynth} terms according to their scores for each state using the terms "default mode" (DM), "salience" (S), "executive" (E), "sensorimotor" (SE), "visual" (V) and "auditory" (A) contrasted against the expected score according to the HMM state transition probabilities. Each plot in the grid shows the score for each term associated with a state activity map, plotted against the mean expected score (under the HMM transition probability $P$) of the next forward (FOR) timestep. The comparison between state score and expected score forward in time demonstrates spatiotemporal relationships between the states (layers) of the network. \label{suppfig:neurofor}}
\end{sidewaysfigure}
\clearpage
\section{Validation with Synthetic Data \label{supp:val}}
This section details the validation of key features of the HMGM selection and analysis framework using synthetic (simulated data). Where possible, similar simulation approaches are used in order to make the simulated pipeline as coherent and true to the data as possible while taking into account computational and practical constraints. 
\subsection{Parallel Analysis Experiments \label{supp:pasim}}
\begin{figure}[!h]
	\centering
	\includegraphics[width=\textwidth]{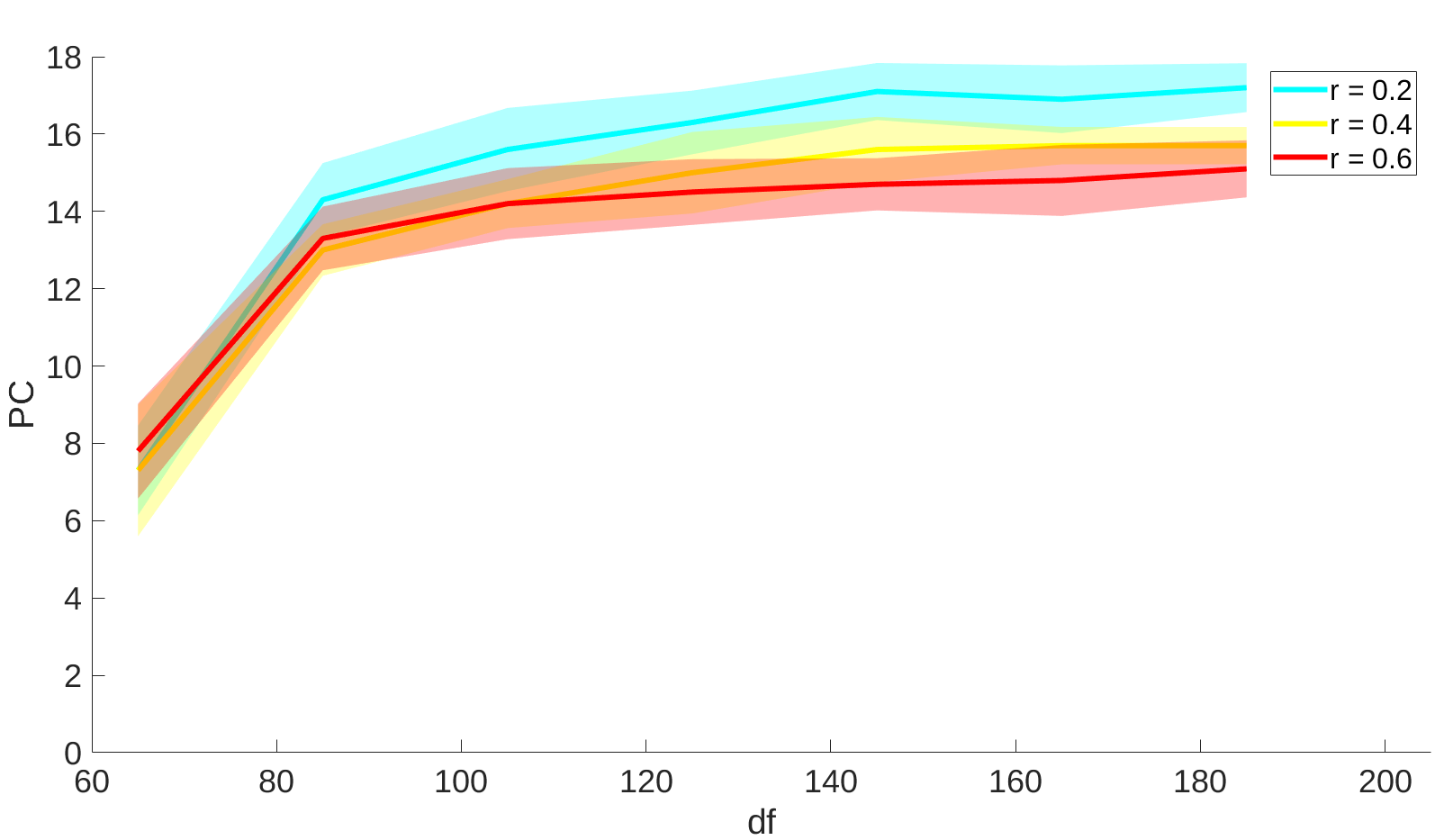}
	\caption{This figure shows the number of Principal Components (PCs) recommended by the Parallel Analysis (PA) method for a small data set of synthetic brain activity data. The colours show the level of intracommunity correlation within the synthetic brain state (based on actual observed community membership). The shading shows the standard error in PCs while the colour indicates the intracommunity correlation in activity. \label{suppfig:pa}}
\end{figure}

We use PCA to reduce the computational complexity and noise in the modelling process. In order to determine whether Parallel Analysis (see Section \ref{supp:pa}) could be used to obtain reasonable and consistent estimates of the embedding dimension (number of PCs) even when variability is high, we performed experiments on synthetic data simulating functional activity from $D=63$ brain regions (the same dimensionality as the real data). Brain activity is simulated from a state $s\in\mathcal{S}$ with multivariate normal observation assumed to have state mean activity $\mu(s)$ and noisy covariance matrix $\Sigma(s)$. These number of states  is the same as the observed $K=27$ state model. 
\\*[4pt]
In this simulation, communities in state $s$ are embedded into a noisy covariance matrix, $\Sigma(s)$, as cliques with 
correlation $r$ (this differs from the variable relationships between brain regions in the same community seen in the observed model). Each community, $C$, is a member of partition $\mathcal{C}_s$. These brain region communities have identical membership to the actual communities in the observed model. This helps to provide clique communities of variable sizes that are consistent with the observed model. In order to account for community structure as well as noise and intersubject variability, the state covariance matrix is generated from an Inverse Wishart distribution with scale matrix $\Psi(s,r)$,
\begin{align*}
\Psi(s,r)_{i,j} = \begin{cases}
1 & \textrm{if }i=j\\
r & \textrm{if } \delta_C(i,j)\\
0 & \textrm{otherwise}.
\end{cases},
\end{align*}
and degrees of freedom $\nu\ge D-1$ which represents the amount of variability in $\Sigma(s)$. The degrees of freedom are low when reflecting noisy data with a lot of inter- and intrasubject variability and high when there is assumed to be little noise in $\Sigma(s)$. We explore the specific case $r>0$ for parametric simplicity. In practice, the particular realisation of the state intracommunity correlation may be positive or negative.
\\*[4pt]
The simulation of dynamic brain activity for a given number of degrees of freedom ($\nu$) and intracommunity correlation $r$ works as follows:
\begin{enumerate}
	\item The state covariance $\Sigma(s)$ matrices are generated from an $IW(\Psi(s,r),\nu)$ distribution with degrees of freedom $\nu$, community membership coming from Louvain community detection as performed on the observed model (with $\gamma=2$), and intracommunity correlation $r$. In addition, state mean activity $\mu(s)$ is generated using a multivariate normal distribution with zero mean and unit variance.
	\item Then $n=20$ samples are generated from the state.
	\item Step 1 and 2 are repeated $t=100$ times to generate a sample of size $N=2000$ which is then mean subtracted and divided by the standard deviation for each dimension respectively to produce a dataset $X$ with zero mean and unit variance.
	\item Parallel Analysis is performed on the sample $X$.
	\item This process is repeated $R=10$ times for a particular parametrisation to generate an empirical distribution of PCs suggested by parallel analysis.
\end{enumerate}
The result of generating samples for different values of $r$ (degree of correlation between regions) is shown in Figure \ref{suppfig:pa} over a range of noise in the state covariance matrix (represented by different $\nu$). When the noise in state covariance is high (i.e. $\nu$ is high) then the number of recommended dimensions is low (suggesting little signal above noise in the data), however as the noise decreases the number of PCs recommended by Parallel Analysis increases to capture more of the signal. The method behaves similarly no matter the intracommunity correlation strength, although for higher values of $r$ the number of PCs required at low noise decreases slightly. This could reflect that the high level of correlation results in fewer PCs needed to represent the data.
\subsection{Clique Community Recovery Experiments \label{supp:cr}}
\begin{figure}[!h]
	\centering
	\includegraphics[width=\textwidth]{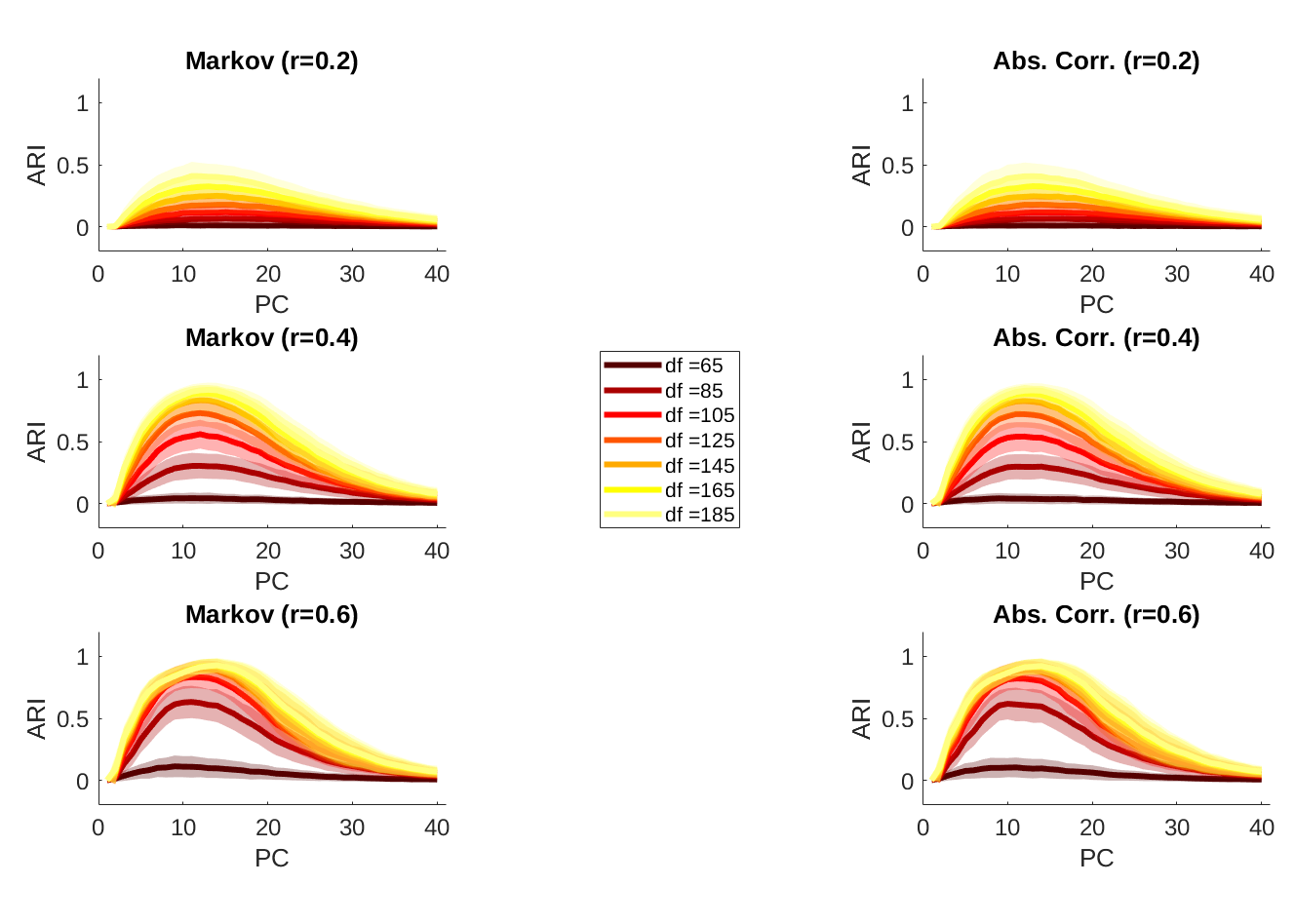}
	\caption{This figure compares the performance of the Louvain community detection algorithm on clique-recovery tasks for two different methods of constructing graph models from simulated state covariance matrices of size $D=63$. The first method is the directed Markov Information matrix method (left) and the second is the method of undirected graph construction by absolute correlation (right). ARI scores for a fixed number of degrees of freedom (df), denoted $\nu$, are plotted across a variable number of principal components (linear embedding dimension), with higher $\nu$ corresponding to reduced noise in the covariance matrix. Performance is almost identical for both methods with performance improving as either the clique strength or degrees of freedom increases. The relationship between linear embedding dimension (PC) and model performance is not monotonic but rather, improves as more signal-rich components are included, deteriorating when higher noisy components are included. \label{suppfig:sims}}
\end{figure}

In order to determine whether state Markov Information matrices (see Equation \eqref{markovFC} of the main text) could reasonably be used to recover the spatial community structure of a state, we tested this method on a clique community recovery task in which clique communities are embedded in a noisy covariance matrix generated as in Section \ref{supp:pasim}. We then compare the performance to a simple undirected method using absolute correlation to generate an graph from a covariance matrix. We also examine the effects of model parameters on clique recovery performance, notably the degrees of freedom $\nu$, the within-clique correlation $r$ and the number of principal components used to reduce the matrix dimension (as in Section \ref{prep} of the main text).
\\*[4pt]
Community detection algorithm performance was computed using the Adjusted Rand Index (ARI) \cite{rand1971objective}, which measures the similarity between the true partition $\mathcal{C}$ and that calculated by the Louvain algorithm. A moderate ARI score is 0.6 or above. 
\\*[4pt]
In our model preprocessing procedure, data is first dimensionally reduced to reduce noise and complexity. We thus perform the same transformation on the sampled covariance matrix as in Equation \ref{eq:back} of the main text. In order to approximate real partitioning, true partitions were sampled from the set of $K=27$ partitions determined from data in the main text and 1000 $D\times D$ matrix realisations of the inverse Wishart distribution were generated for each setting of the simulation parameters: degrees of freedom, PCs and within-clique correlation ($r$). Figure \ref{suppfig:sims} shows the result of these experiments for both methods. The results show that for moderate levels of within-clique correlation, both graph construction methods are able to recover true community activity for a reasonably large range of parameters. Notably, when the number of principal components is either too high (including too much noise) or too low (removing too much signal), community detection performance suffers, underscoring the importance of reasonably choosing this parameter.
\\*[4pt]
The results suggests that the optimum number of PCs to recover the community structure embedded in a covariance matrix (depending somewhat on the amount of noise) is between 10 and 20. This is in agreement with the results of Parallel Analysis simulation (see Figure \ref{supp:pa}) as even when noise is relatively high ($\nu\ge 105$), the number of recommended PCs is above 12. If the HMM training procedure is able to recover a moderately accurate representation of the covariance matrix (with some allowance for noise), and the within community correlation is strong ($r\ge 0.4$), our simulation indicates that parallel analysis is able to recover the community structure with relative accuracy (having a moderate ARI).
\subsection{Hidden Markov Model Selection Experiments \label{supp:hmmsim}}
\begin{sidewaysfigure}[!h]
	\centering
	\includegraphics[width=\textwidth]{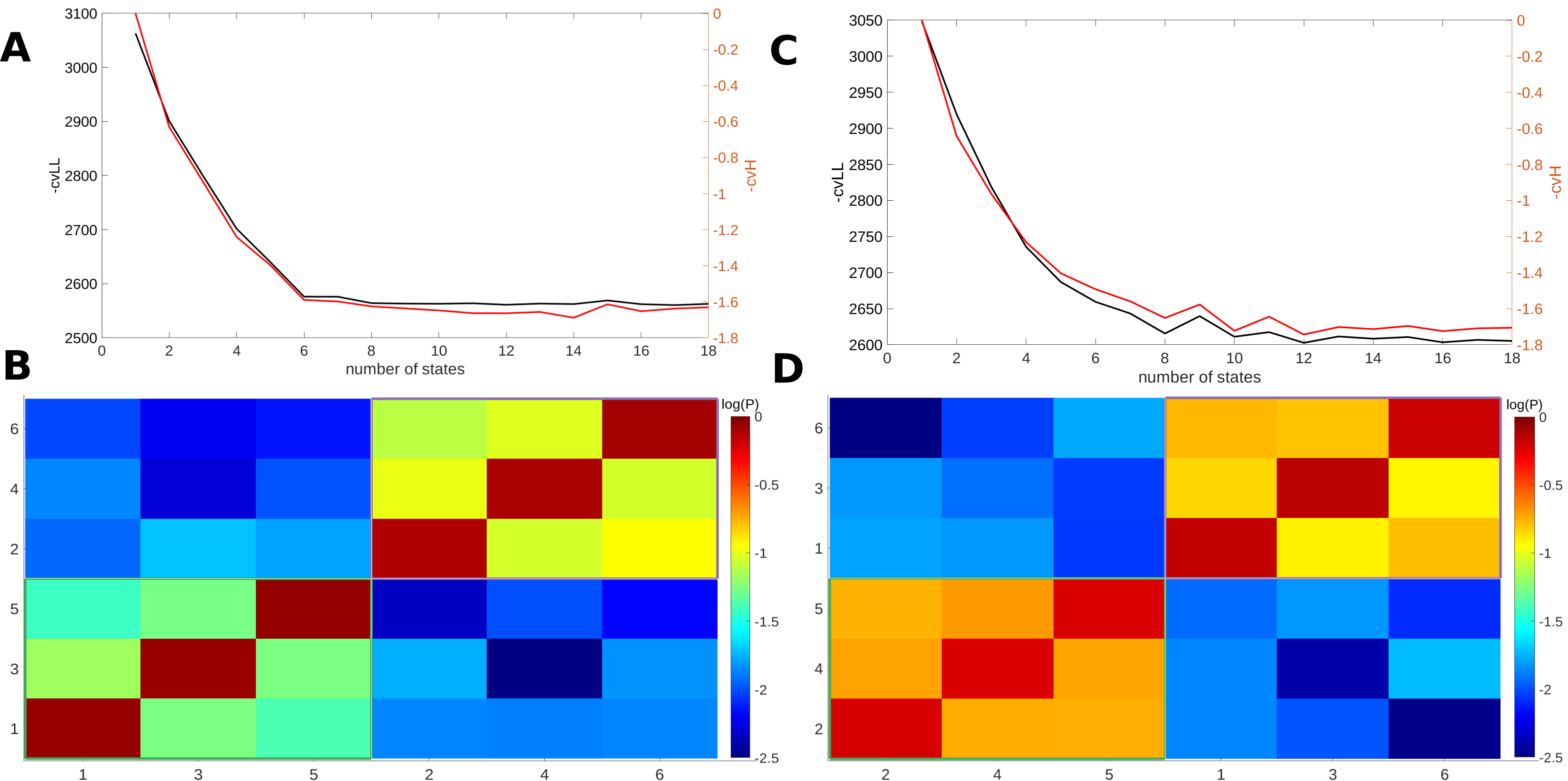}
	\caption{Model selection and temporal community detection based on synthetic data simulated from $N=15$ subjects with subject-specific noise. (A) Shows the results of model selection using the cross-validated log-likelihood (black, left axis) and cross-validated entropy (red, right axis). (B) Shows the log-transition matrix (colorbar with log-probability of transition) for the $c=0.05$ model with lines dividing the two communities detected using the Louvain method with $\gamma=0.08$ (determined by minimum Variation of Information). The axis labels are the states. The exact state label is arbitrary and so should not be expected to match the original numerical labels. (C) This figure shows the model selection results for for the higher intracommunity edge strength model with $c=0.15$. (D) Shows the temporal community detection results for the final model with again $\gamma=0.08$. In both cases it is clear that the method recovers the correct community structure. \label{suppfig:hmmsim}}
\end{sidewaysfigure}

Finally, we test the ability of the cross validated maximum entropy model selection procedure to select the correct number of HMM states from synthetic data that has already been dimensionally reduced to $d=9$ dimensions. We apply the same additional constraint as in Section \ref{msel_res}, that the state must be present in at least 25\% of subjects to be included. In addition to testing whether the selection criterion can identify the correct number of states, we also test whether the model identified and trained on this data can recover the temporal community structure of the states.
\\*[4pt]
Data from $N=15$ subjects was generated over $T=200$ time points. The data was generated from an HMM with $K=6$ states and $d$ MVN observations that includes subject-specific noise in the covariance matrix. The number of states is considerably less than for the observed model but this was done due to computational constraints on simulating and running multiple models.
\\*[4pt] 
Each state covariance matrix is assumed $\Sigma(s)\sim IW(\mathbf{I},40)$ distributed (chosen to be roughly $\nu \approx d+30$ for reasonable hypothetical recovery if community structure were present), where $\mathbf{I}$ is the identity matrix, with mean $\mu(s)\sim MVN(\mathbf{0},\mathbf{I})$. The state transition matrix $P(c)$ was chosen so that given correctly selected $c>0$ the matrix can be potentially partitioned into two communities, a strongly connected community and a weakly connected community, each with equal membership, so that $P(c)$ equals
\begin{align*}
\left(\begin{array}{ccc|ccc}
0.97 - 2c & c & c & 0.01 & 0.01 & 0.01\\
c & 0.97 - 2c & c & 0.01 & 0.01 & 0.01\\
c & c & 0.97 - 2c & 0.01 & 0.01 & 0.01\\\hline
0.01 & 0.01 & 0.01 & 0.87 - 2c & c+0.05 & c+0.05\\
0.01 & 0.01 & 0.01 & c+0.05 & 0.87 - 2c & c+0.05\\
0.01 & 0.01 & 0.01 & c+0.05 & c+0.05 & 0.87 - 2c\\
\end{array}\right),
\end{align*}
where the lines indicate the two temporal communities with the first being the weaker community. As in the observed model for wakefulness, the probability of self-transition is high in general with a relatively low probability of leaving a given community of $10^{-2}$.
\\*[4pt]
In order to incorporate subject specific differences in state activity, observations for subject $i$ are sampled from a slightly perturbed model with $\Sigma(s,i)=\Sigma(s)+\epsilon\hat\Sigma$ where $\hat \Sigma$ is sampled from the same distribution as $\Sigma(s)$ and similarly $\mu(s,i)=\mu(s)+\epsilon\hat\mu$ where $\hat \mu$ is sampled from the same distribution as $\mu(s)$. Here we choose $\epsilon=0.01$.
\\*[4pt]
Figure \ref{suppfig:hmmsim}A shows the results of model selection and HMM fitting when the temporal community coupling parameter is set to $c=0.05$. As in the observed model (see Section \ref{msel_res}), there is a clear relationship between cross-validated log-likelihood maximisation and entropy maximisation. The negative cross-validated entropy appears to decrease near monotonically with the number of states, as does the likelihood. The resulting minimum entropy model has $K=14$ initial states, 8 of these were removed due to not being present in enough subjects and providing a final model with 6 states as expected. It is also clear from Figure \ref{suppfig:hmmsim}B that the method recovers the underlying community structure of the network even when the preference for intracommunity transition over intercommunity transition is relatively weak. The difference in intracommunity transition probability are also evident from the trained model. Similar results are seen when $c=0.15$, in Figure \ref{suppfig:hmmsim}C. In this case initially $K=12$, but pruning non-general states again gives 6 states in the final model with the expected community structure. The fact that the number of states is overestimated in both cases shows the importance of the pruning step in determining the final model. In addition, the optimal resolution, $\gamma$, according to the Variation of Information was found to be the same for both models with $\gamma=0.08$.
\\*[4pt]
In both models it is clear that the entropy increases with additional states but the increases begin to slow down after the true number of states is exceeded. It thus may be beneficial to consider not only the absolute global minimum but also the first local minimum when assessing potentially viable models for exploration.
\clearpage  
\clearpage
\end{document}